\documentclass[12pt]{article}
\pdfoutput=1
\usepackage{amsmath}
\usepackage{amsfonts}
\usepackage{float}
\usepackage{url}
\usepackage{hyperref}
\usepackage{mathtools}
\usepackage{bbold}
\usepackage{slashed}
\usepackage{simplewick}
\usepackage{tikz}
\usepackage{graphicx}
\usepackage{epstopdf}
\usepackage{subfigure}
\usepackage[labelfont=bf,font={small}]{caption}
\usepackage{pgfplots}
\usepackage{amssymb}
\usepackage{color}
\usepackage{mathrsfs}
\usepackage{fancybox}
\usepackage{lipsum}
\usepackage{eurosym}
\usepackage{tcolorbox}
\usepackage{physics}
\usepackage{tensor}
\usepackage{empheq}

\numberwithin{equation}{section}
\newcommand{\nn}{\nonumber}
\definecolor{airforceblue}{rgb}{0.36, 0.54, 0.66}
\newcommand{\beq}{\begin{equation}}
\newcommand{\eeq}{\end{equation}}

\newcommand{\sltr}{\text{SL}(2,\text{R})}
\newcommand{\dpi}{\mathcal{D}}

\newcommand{\mo}{\mathcal{O}}
\newcommand{\mm}{\mathcal{M}}

\newcommand{\average}[1]{\left\langle #1 \right\rangle}

\begin{document}\begin{titlepage}

\setcounter{page}{1} \baselineskip=15.5pt \thispagestyle{empty}

%\begin{flushright}
%hep-th/13mmnnn\\
%\end{flushright}
\vfil
${}$
%\vspace{0cm}
\begin{center}
{\Large \bf Searching for butterflies in dS JT gravity}
\vskip 1cm

\textbf{Andreas Blommaert}

\vspace{0.5cm}

{\sl Stanford Institute for Theoretical Physics,\\ Stanford University, Stanford, CA 94305 \\}

\vspace{0.2cm}

\medskip

\end{center}

\vskip 1cm

\begin{center}
{\bf Abstract}
\end{center}
We investigate out of time ordered correlators in the bulk of dS JT gravity, using Schwarzian perturbation theory, and propose that these out of time ordered correlators are encoded on the second sheet of the gravitational path integral, different sheets corresponding to different gravitational operator orderings. Implementing this in practice, we establish maximal chaos, in agreement with shockwave intuition.

\end{titlepage}

\newpage
\tableofcontents
\vspace{0.5cm}
\noindent\makebox[\linewidth]{\rule{\textwidth}{0.4pt}}
\vspace{1cm}
%\newpage

\addtolength{\abovedisplayskip}{.5mm}
\addtolength{\belowdisplayskip}{.5mm}

\def\plus{\raisebox{.5pt}{\tiny$+$\smpc}}

\addtolength{\parskip}{.6mm}
\def\spc{\hspace{1pt}}

\def\nspc{{\hspace{-2pt}}}
\def\ff{\rm\smpc f\smpc} 
\def\fff{\mbox{Y}}
\def\ww{{\rm w}}
\def\smpc{{\hspace{.5pt}}}

\def\zz{{\spc \rm z}}
\def\xx{{\rm x\smpc}}
\def\xxi{\mbox{\footnotesize \spc $\xi$}}
\def\jj{{\rm j}}
\addtolength{\baselineskip}{-.1mm}

\renewcommand{\Large}{\large}

\setcounter{tocdepth}{2}
\addtolength{\baselineskip}{0mm}
\addtolength{\parskip}{-0.6mm}
\addtolength{\abovedisplayskip}{1mm}
\addtolength{\belowdisplayskip}{1mm}

\setcounter{footnote}{0}
\section{Introduction}\label{sec:intro}

Many exciting developments over the last decade have emphasized the key role of quantum chaos in quantum gravity. For example we now know that black holes in AdS are maximally chaotic quantum systems \cite{sekino2008fast, shenker2014black,shenker2014multiple, roberts2015localized, Roberts:2014ifa,jackson2015conformal, shenker2015stringy, maldacena2016bound}. Another important milestone is the realization that the SYK model \cite{Sachdev:1992fk,kitaev} has a dual description as AdS$_2$ quantum gravity \cite{maldacena2016remarks, Jensen:2016pah, Maldacena:2016upp, Engelsoy:2016xyb, kitaev2018soft}. The discovery of this maximally chaotic quantum system spurred new developments in AdS$_2$ quantum gravity  \cite{maldacena2017diving, kourkoulou2017pure, maldacena2018eternal, lin2019symmetries}, and its relation to random matrices \cite{cotler2017black, saad2018semiclassical, Saad:2019lba, Stanford:2019vob,saad2019late,Blommaert:2019wfy,Marolf:2020xie,Blommaert:2020seb,Mertens:2020hbs,Maxfield:2020ale,Witten:2020wvy}. This tightens the connection between gravity and quantum chaos as random matrices are an effective description of generic chaotic quantum systems \cite{mehta2004random,haake1991quantum}.

Our understanding of quantum gravity in dS is lagging behind significantly as compared to our current knowledge of quantum gravity in AdS. This is largely because the holographic correspondence in dS quantum gravity is much less intuitive as compared to its AdS cousin. There are several candidates for a holographic correspondence in dS \cite{Witten:2001kn, Strominger:2001pn, maldacena2010vacuum,Dong:2010pm, Gorbenko:2018oov} but it might also be that none of them makes sense \cite{Dyson:2002nt}. More broadly it is unclear what is the correct computational framework for quantum gravity in dS \cite{Witten:2001kn, Strominger:2001pn, maldacena2010vacuum, freivogel2006holographic, Anninos:2017eib, Anninos:2019nib, Maldacena:2019cbz, Cotler:2019nbi, cotler2019emergent}. 

What we need to make progress is a tractable model of quantum gravity in dS.

Enter JT gravity. This theory of dilaton gravity has been popular to study AdS$_2$ quantum gravity. Signatures of quantum chaos in JT gravity were investigated in \cite{Maldacena:2016upp,Mertens:2017mtv,Lam:2018pvp,saad2018semiclassical,saad2019late,Saad:2019lba,Blommaert:2020yeo,Blommaert:2020seb,Kapec:2019ecr}. Here we will investigate signatures of quantum chaos in the dS$_2$ version of JT gravity \cite{Maldacena:2019cbz, Cotler:2019nbi, cotler2019emergent}. The emergence of random matrices in dS JT gravity was investigated in \cite{Saad:2019lba,Maldacena:2019cbz,Cotler:2019nbi}. Another hallmark feature of chaotic quantum systems is exponential sensitivity to changes in initial conditions, which should be probed by computing out of time ordered correlators \cite{shenker2014black,shenker2014multiple, roberts2015localized, Roberts:2014ifa,jackson2015conformal, shenker2015stringy, maldacena2016bound}. It is suspected that dS horizons are maximally chaotic, just like black holes in AdS \cite{sekino2008fast,susskind2011addendum}. This means that Lyapunov exponents of out of time ordered correlators in dS are suspected to saturate the bound on chaos $\lambda_L=2\pi/\beta$ \cite{maldacena2016bound}.

The notion of such a bound on chaos is not guaranteed in the context of dS quantum gravity. The bound, due to Maldacena, Shenker and Stanford \cite{maldacena2016bound} applies to specific quantum systems and in particular holds for out of time ordered correlators computed in the CFT dual description of quantum gravity in AdS. In case of dS quantum gravity one might hope to similarly appeal to a dual description to derive a bound on chaos. But known boundary duals to theories of dS quantum gravity are Euclidean field theories, for which the proof of \cite{maldacena2016bound} does not apply. Furthermore we are asking strange questions of this dual theory, as out of time ordered correlators in dS are bulk correlators. How is dynamical chaos encoded in this nondynamical boundary dual?

That out of time ordered correlators in dS are bulk observables is conceptually perhaps the biggest hurdle to overcome in order to do meaningful calculations. This is because the diff redundancy of quantum gravity forces us to define bulk operators with a gravitational dressing \cite{Donnelly:2016rvo,Donnelly:2015hta,Engelhardt:2016wgb,Lewkowycz:2016ukf}. Another related issue is how to implement out of time ordering for bulk observables. How does one practically go about folding the bulk time contour?

There is recent evidence obtained by means of semiclassical shockwave computations in dS$_3$ supporting the claim that dS horizons are maximally chaotic \cite{Aalsma:2020aib}. However the analysis of \cite{Aalsma:2020aib} does not take into account the requirement to define bulk operator insertions in a diff invariant manner in quantum gravity, and therefore cannot be completely trusted.

The analysis in AdS$_3$ \cite{shenker2014black} does not suffer from this, because they use boundary operators.

In this work, we present evidence for maximal chaos in out of time ordered correlators in dS JT gravity. We do so by adapting methods to define, and compute diff invariant bulk correlators in AdS JT gravity \cite{blommaert2019clocks}. One important part of this analysis is a proposal for how to fold the time contour in the bulk, which boils down to specifying an analytic continuation. Working under this assumption, we find $\lambda_L=2\pi/\beta$ and we match to a shockwave analysis.
\\~\\
This work is organized as follows.

In \textbf{section \ref{sec:BulkMatter}} we introduce the conceptual framework for dS$_2$ quantum gravity with which we will be working. We construct and compute diff invariant bulk correlators, and interpret the result in terms of bulk operator reconstruction, following \cite{blommaert2019clocks,Hamilton:2005ju,Hamilton:2006az,Kabat:2011rz,Kabat:2017mun,xiao2014holographic}.

In \textbf{section \ref{sec:shockandscramble}} we discuss and address the conceptual obstacle of implementing folding of the bulk time contour. Using Schwarzian perturbation theory \cite{Maldacena:2016upp,Sarosi:2017ykf}, we establish maximal chaos, given some prescription for implementing bulk gravitational out of time ordering.

In \textbf{section \ref{sect:ads}} as an aside, we briefly discuss bulk out of time ordered correlators in AdS$_2$.

In \textbf{section \ref{sec:discussion}} we compare with shockwave intuition and provide closing remarks.
%%%%%%%%%%%%%%%%%%%%%%%%%%%%%%%%%%%%%%%%%%%%%%%%%%

%%%%%%%%%%%%%%%%%%%%%%%%%%%%%%%%%%%%%%%%%%%%%%

%%%%%%%%%%%
% SECTION %
%%%%%%%%%%%
\section{Framework}
\label{sec:BulkMatter}
The Lorentzian action for dS JT gravity is \cite{Maldacena:2019cbz, Cotler:2019nbi}
\begin{equation}
\label{E:action1}
S_\text{gravity} = S_0 \,\chi + \frac{1}{16 \pi G} \int d^2 x \sqrt{-g} \,\phi \, (R-2) - \frac{1}{8\pi G} \int_{\partial} dx \sqrt{h} \,\phi\, (K-1) \,.
\end{equation}
We will consider path integrating over spacetimes with one circular spacelike boundary. Let us impose the usual fixed length boundary conditions \cite{Maldacena:2016upp}, where we introduce a $2\pi$ periodic coordinate $u$, which is proportional to the proper length along this boundary, as
\begin{equation}
    ds\rvert_{\partial \mm}=\frac{\ell}{2\pi\varepsilon}\,du\quad,\quad \varphi\rvert_{\partial \mm}=\frac{1}{\varepsilon}\quad,\quad \varepsilon\to 0\,.\label{bc}
\end{equation}
Here $\ell$ is the renormalized length of this asymptotic boundary.\footnote{We reserve $\beta$ for the inverse dS temperature which depends on the cosmological constant as $\beta = 2\pi/\sqrt{\Lambda}$.} Path integrating over bulk values of the dilaton $\phi$ localizes on metrics with constant positive curvature $R+2=0$. 

The simplest geometries with one spacelike fixed length boundary are Hartle-Hawking geometries, with a wiggly boundary near future infinity \cite{Maldacena:2016upp,Engelsoy:2016xyb}
\begin{equation}
    \raisebox{-10mm}{\includegraphics[width=39mm]{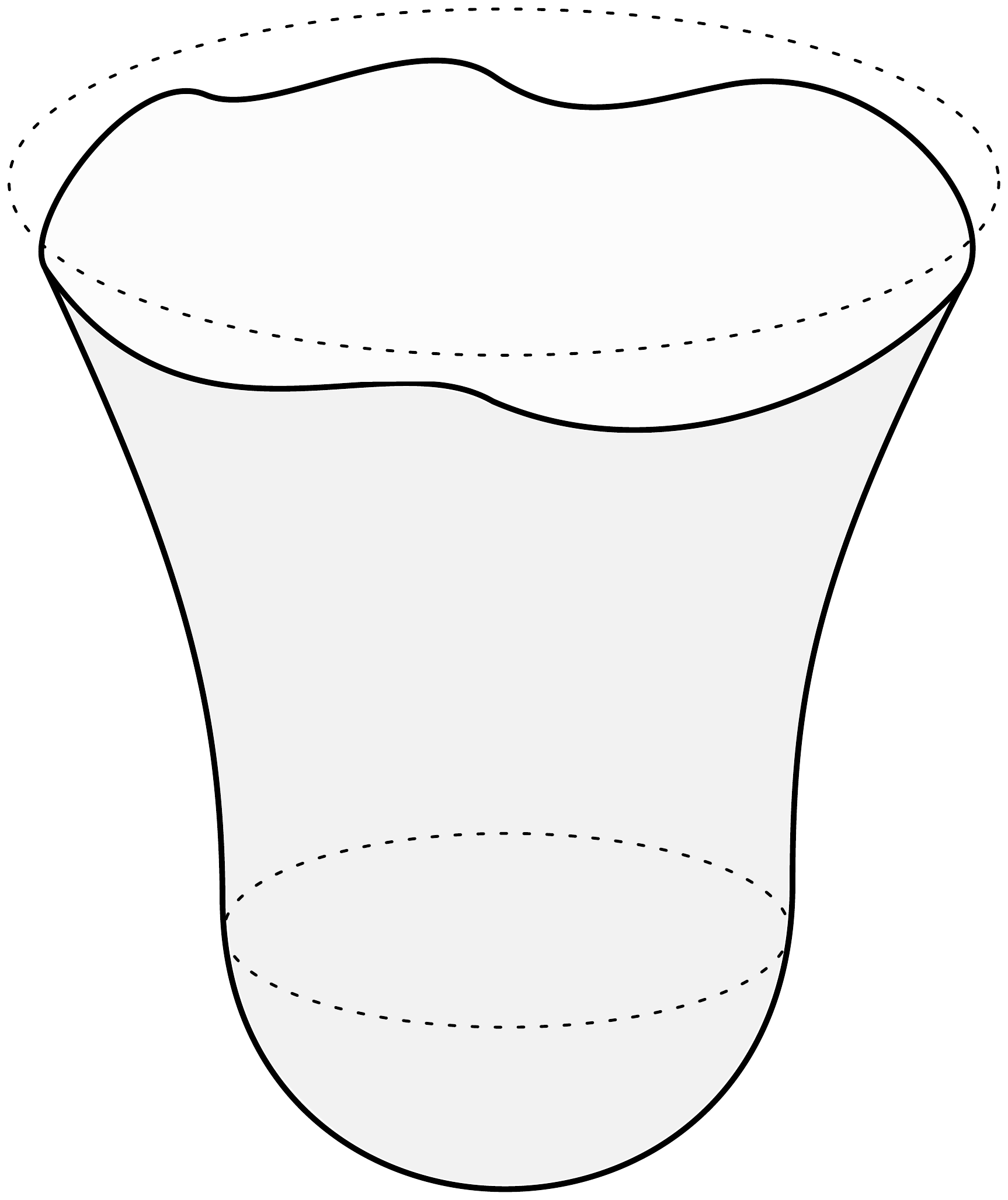}}\label{hartle}
\end{equation}
These complex geometries can be thought of as gluing some Lorentzian expanding geometry
\begin{equation}
ds^2 =\cosh^2t\, d\theta^2-dt^2\,, \label{expan}
\end{equation}
to a Euclidean half sphere at $t=0$. Equivalently, by deforming the complex time contour, we may view these geometries as analytic continuations of the AdS$_2$ disk, with fixed length boundary conditions \cite{Maldacena:2016upp,Maldacena:2019cbz,Cotler:2019nbi}. For our purposes it is convenient to transform to conformal coordinates which cover the same patch
\begin{equation}
\label{E:conformalcoords1}
ds^2 = \frac{d\theta^2-d T^2}{\cos^2 T}\,,
\end{equation}
where $T$ approaches $\pi/2$ at the asymptotic boundary. 

The location of the wiggly boundary near future infinity is fixed entirely by specifying the function $\theta=f(u)$. The action then reduces to the usual Schwarzian action \cite{Jensen:2016pah,Maldacena:2016upp,Engelsoy:2016xyb,Maldacena:2019cbz,Cotler:2019nbi}
\begin{equation}
\label{E:action2}
S_\text{gravity} = -i S_0 + \frac{1}{4 G \ell} \int_0^{2\pi} du \,\text{Sch}(F,u)\quad,\quad F(u)=\tan \frac{f(u)}{2}\,.
\end{equation}
The corresponding path integral is identical to the Euclidean AdS$_2$ disk path integral with replacement $\beta=-i\ell$. We integrate over $2\pi$ periodic fields $f(u)$ modulo redundant global $\sltr$ transformations, and with the standard measure for the Schwarzian path integral \cite{Stanford:2017thb}. Gauge fixing the global redundancy in the usual way, the classical saddle is $f(u)=u$.

The gravitational path integral with these boundary conditions gets contributions from higher genus topologies. These complex spacetimes are analytic continuations of the AdS$_2$ disk with handles, which asymptote to the expanding geometries \eqref{expan} \cite{Saad:2019lba,Maldacena:2019cbz,Cotler:2019nbi}. These contributions are suppressed by genus, and can be ignored at the scrambling time scale \cite{Saad:2019lba}. Therefore we consider only genus zero henceforth.
%%%%%%%%%%%%%%%%%%%%%%%%%%%%%%%%%%%%%%%%%%%
\subsection{Invariant bulk points}\label{sect:cr}
We now consider free minimally coupled massive scalar fields
\begin{equation}
\label{fieldact}
S_{\text{matter}} = -\frac{1}{2} \int d^2 x\sqrt{-g}\, g^{\mu \nu}\partial_\mu \varphi\, \partial_\nu \varphi + m^2 \varphi^2\,,
\end{equation}
and focus on continuous series  representations
\begin{equation}
    m^2=\Delta(1-\Delta)>\frac{1}{4}\,.\label{range}
\end{equation}
We are path integrating over Hartle-Hawking geometries, therefore the matter fields will be most naturally prepared in the Bunch-Davies vacuum at $t = 0$.

With the eye on out of time ordered correlators, we want to study correlation functions of matter fields inserted at finite Lorentzian times in the bulk. To this end we must decide on a geometric construction, which specifies the locations of bulk operator insertions. Indeed, we will be integrating over metrics $g_f$ which depend on $f\,$, hence we require a coordinate independent way to define a bulk point common to each $g_f\,$ \cite{Donnelly:2016rvo,Donnelly:2015hta,Engelhardt:2016wgb,Lewkowycz:2016ukf}.  The geometries $g_f$ all share the boundary conditions \eqref{bc} so that we can leverage the boundary to geometrically define bulk points \cite{Donnelly:2015hta,Almheiri:2017fbd,Lewkowycz:2016ukf,Chen:2017dnl,Chen:2018qzm,Verlinde:2015qfa,Nakayama:2015mva,Goto:2016wme,Engelhardt:2016wgb,blommaert2019clocks,Mertens:2019bvy,Blommaert:2020yeo,Blommaert:2020seb}. The point is that obviously the proper length coordinate along the boundary is a geometric, or diff invariant coordinate.

Our construction closely follows that of \cite{blommaert2019clocks}. Parameterizing the boundary as $\theta=f(u)$, we choose points $u_1$ and $u_2$ that are separated by a fixed proper length along the boundary curve. Imagine shooting a lightray back in time to the right from $u_1$, and shooting another lightray back in time to the left from $u_2$. The intersection of these two rays defines a unique point in the bulk, which we denote by $(u_1,u_2)$. The definition of this point is geometric, and therefore independent of bulk coordinates, so it makes sense in a path integral over metrics.

In conformal coordinates, and on the saddle $f(u)=u$, the point $(u_1,u_2)$ corresponds to
\begin{equation}
T = \frac{\pi}{2} - \frac{1}{2}(u_2 - u_1 \,\,(\text{mod }2\pi) )\quad, \quad \theta = \frac{1}{2}(u_1 + u_2)\,.
\end{equation}
For more general reparameterizations $f(u)$, the point $(u_1, u_2)$ corresponds to \cite{blommaert2019clocks}
\begin{equation}
T = \frac{\pi}{2} - \frac{1}{2}(f(u_2) - f(u_1) \,\,(\text{mod }2\pi) )\quad,\quad\theta = \frac{1}{2}(f(u_1) + f(u_2))\,.\label{210}
\end{equation}
So in rigid conformal coordinates \eqref{E:conformalcoords1}, the point $(u_1,u_2)$ is fuzzy.

With this definition of bulk points we can compute bulk observables like
\begin{equation}
\label{E:W1}
\frac{1}{Z_{\text{total}}} \int [\mathcal{D}f]\,e^{i S_\text{gravity}[f]} \int [\mathcal{D}\varphi] \, e^{i S_{\text{matter}}[\varphi, \,f]}\, \varphi(u_1, u_2) \varphi(u_1', u_2')\,.
\end{equation}
This means little, unless we specify boundary conditions and ordering ambiguities. We can visualize the framework within which we will be doing computations as
\begin{equation}
    \quad \raisebox{-20mm}{\includegraphics[width=39mm]{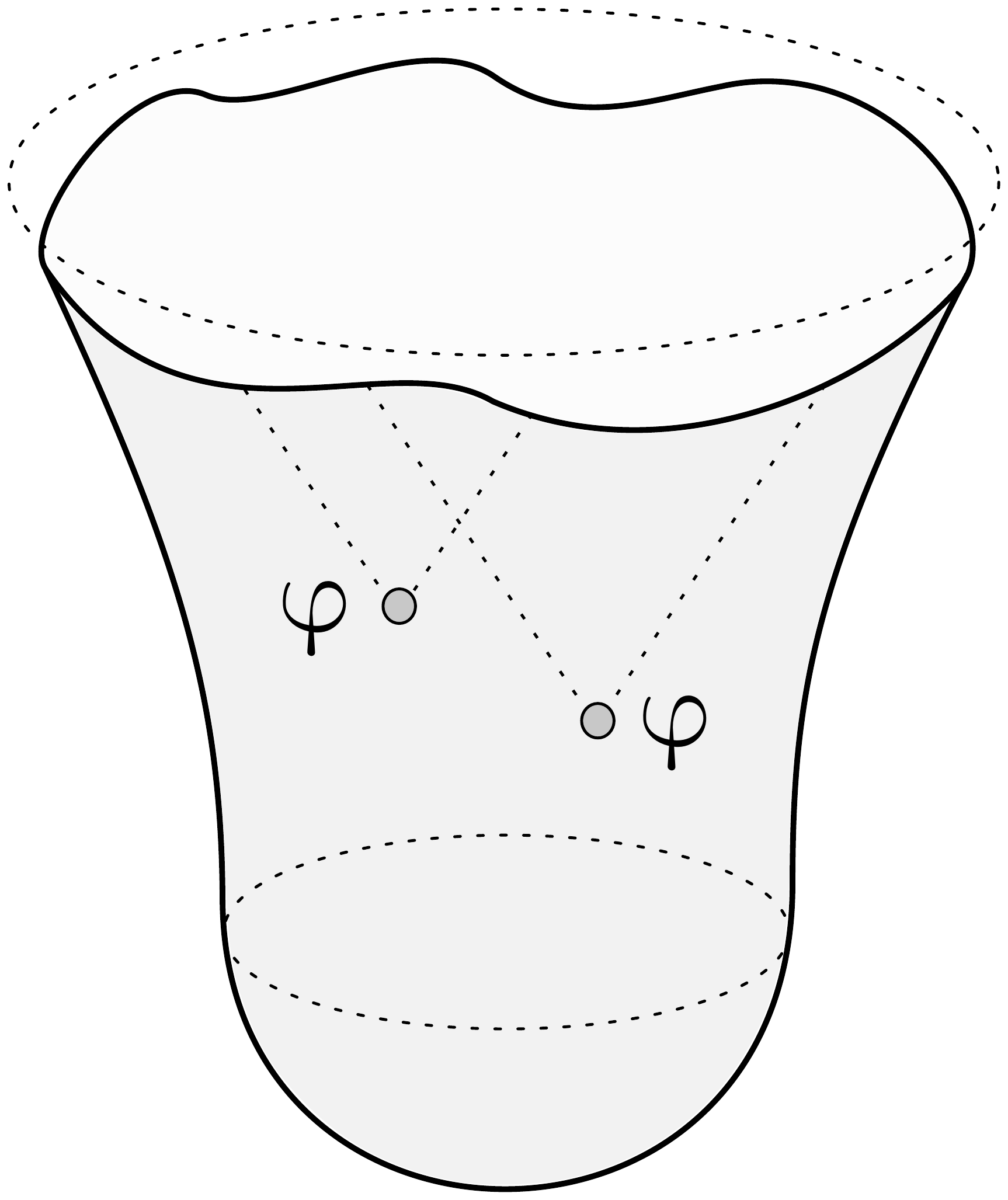}}\quad\label{forback}
\end{equation}
In path integral language, we choose the past boundary conditions on the combined gravity and matter system to prepare the Hartle-Hawking and Bunch-Davies states at $T=0$. The future boundary conditions for the gravitational field are \eqref{bc}, this defines an $\ket{\ell}$ state \cite{cotler2019emergent}. We choose the future boundary conditions on the matter fields to correspond with infinite time evolution of the Bunch-Davies state. With these boundary conditions we then compute the Feynman (or time ordered) propagator of the matter fields.

In quantum mechanical notation we write
\begin{equation}
    \bra{\text{HH}}\otimes \bra{\text{BD}}\mathcal{T}(\varphi(u_1, u_2) \varphi(u_1', u_2'))\ket{\ell}\otimes \ket{\text{BD}}\label{bdbd}
\end{equation}
Time evolution operators are left implicit, to lighten notation. This represents the transition amplitude from an initial state $\ket{\text{HH}}\otimes \ket{\text{BD}}$ to an out state $\ket{\ell}\otimes \ket{\text{BD}}$, where the evolution operator is defined to introduce two scalar field operator insertions, at specified points.

This may look like a peculiar observable to compute in a cosmological context, because free time evolution obviously doesn't insert scalar fields at specified points. However we will see (in section \ref{sect:bulkrec}) that we can wield bulk operator reconstruction to rewrite this observable, as some complicated linear combination of wavefunction components
\begin{equation}
    \psi(\ell,\varphi)=\bra{\text{HH}}\otimes \bra{\text{BD}}\,U\,\ket{\ell}\otimes \ket{\varphi}\label{wave}
\end{equation}
where $\ket{\varphi}$ denotes some multi particle state in the matter state space in the infinite future. These amplitudes are the metaobservables advocated by Witten as the objects one naturally computes in cosmology \cite{Witten:2001kn}, our bulk observables are just complicated such metaobservables, linear combinations of these standard wavefunction components.

For clarity, this should be compared with another framework for observables in cosmology \cite{Maldacena:2019cbz}, where infinite time matter correlators are computed schematically as expectation values in the some wavefunction $\psi(\ell,\varphi)$, which in particular includes an integration over final states. Here we compute instead transition amplitudes, or wavefunction \emph{components} for specific and crucially fixed final states.

Confused readers are invited to consider my personal view on frameworks in cosmology. We do not genuinely know which framework is best, or even if that question makes sense, the cautious way to proceed is to take potentially interesting results in any framework seriously.

Moving on to computing these observables, consider the matter path integral for a fixed $g_f$. The Bunch-Davies state is defined with respect to the time coordinate $T$, and so does not know about boundary conditions near the far future. Our construction of diff invariant bulk points however, inserts operators using time coordinate $t$. The reparameterization enters the calculation of the matter correlator for fixed $g_f$, because it fixes the relation between these coordinate systems. Observers with proper time $t$ have different experiences than observers with proper time $T$, this is essentially the Unruh effect.

Consequently we can view the matter correlator computed by $t$ observers in the $T$ vacuum as being computed in a reparameterized gravitational background $g_f$
\begin{equation}
    ds^2_f=-\frac{f'(u)f'(v)}{\sin^2\left(\frac{f(u)-f(v)}{2}\right)}\,du\,dv\,.
\end{equation}
The matter part of \eqref{E:W1} hence computes the two point function in a fixed such background
\begin{equation}
\langle \varphi(u_1, u_2) \varphi(u_1', u_2') \rangle_{f} = \frac{1}{Z_{\text{matter}}}\int [\mathcal{D}\varphi] \, e^{i S_{\text{matter}}[\varphi,f]}\,  \varphi(u_1, u_2) \varphi(u_1', u_2') \,,
\end{equation}
which explicitly becomes \cite{blommaert2019clocks,xiao2014holographic}
\begin{equation}
\langle \varphi(u_1, u_2) \varphi(u_1', u_2') \rangle_{f}  = \,_2F_1(\Delta,1-\Delta,1,z_f)\quad,\quad z_f =\frac{\sin \left(\frac{f(u_2) - f(u_1')}{2}\right) \,\sin\left(\frac{f(u_2') - f(u_1)}{2}\right)}{\sin \left(\frac{f(u_2) - f(u_1)}{2}\right) \sin \left(\frac{f(u_2') - f(u_1')}{2}\right)}\,.\nonumber
\end{equation}
We rescaled the scalar fields to absorb constant prefactors, and leave the regularization for timelike separations that specifies the Feynman propagator, implicit \cite{Einhorn:2002nu}. The reader should not get confused about this regularization. It plays no part in our discussion on out of time ordered correlators, which is constructed to isolate entirely gravitational ordering subtleties.

For clarity we emphasize the bulk matter path integral is computing a sphere correlator, because there are two Bunch-Davies states in \eqref{bdbd}, but the insertion locations of operators on the analytically continued sphere depend on the reparameterization, as in \eqref{210}.

We now consider the gravitational path integral
\begin{equation}
\langle \varphi(u_1, u_2) \varphi(u_1', u_2') \rangle = \frac{1}{Z_{\text{gravity}}} \int [\mathcal{D}f] \, e^{i S_\text{gravity}[f]}\, \langle \varphi(u_1, u_2) \varphi(u_1', u_2') \rangle_{f}\,,\label{Wightman1}
\end{equation}
with boundary conditions as previously specified. This generalizes to higher point functions. In the following section, we compute such path integral over metrics perturbatively around the classical saddle. Exact evaluation of these Schwarzian path integrals is possible, however higher order effects are not required to investigate scrambling, hence we will not pursue this.

We note that we can extract boundary observables from these bulk observables. Consider $m^2<1/4$ here, then for $\delta\ll 1$ and $\delta'\ll 1$ and up to constant field rescalings we obtain
\begin{align}
\langle \varphi(u, u + \delta) \varphi(u', u' + \delta') \rangle_{f} =\delta^{\Delta}\,{\delta'}^{\Delta}\,\frac{f'(u)^\Delta f'(u')^\Delta}{\sin^{2\Delta}\left(\frac{f(u) - f(u')}{2}\right)}
\end{align}
Using the extrapolate dictionary, we extract the usual boundary correlator in dS$_2$ \cite{Maldacena:2019cbz, Cotler:2019nbi}
\begin{equation}
\label{E:bdybdy1}
\langle \mo_\Delta(u) \mo_\Delta(u') \rangle = \frac{1}{Z_{\text{gravity}}} \int [\mathcal{D}f] \, e^{i S_\text{gravity}[f]}\,\frac{f'(u)^\Delta f'(u')^\Delta}{\sin^{2\Delta}\left(\frac{f(u) - f(u')}{2}\right)}\,.
\end{equation}

Finally, note that these formulas are similar to those for bulk matter correlators in AdS$_2$, where one typically focuses on discrete series representations
\begin{equation}
    m^2=\ell(\ell-1)\,.
\end{equation}
The thermal patch of AdS$_2$ is parameterized as
\begin{equation}
    ds^2=\frac{d Z^2-d T^2}{\sinh^2 Z}\,.
\end{equation}
Shooting in lightrays from the boundary, we associate bulk points $(u_1,u_2)$ to \cite{blommaert2019clocks}
\begin{equation}
    Z=\frac{1}{2}(f(u_2)-f(u_1))\quad,\quad T=\frac{1}{2}(f(u_2)+f(u_1))\,.\label{cooads}
\end{equation}
The propagator in a reparameterized metric is
\begin{equation}
    \langle \varphi(u_1, u_2) \varphi(u_1', u_2') \rangle_{f}  =\frac{1}{z_f^\ell}\,_2F_1(\ell,\ell,2\ell,1/z_f)\quad,\quad z_f =\frac{\sinh \left(\frac{f(u_2) - f(u_2')}{2}\right) \,\sinh\left(\frac{f(u_1) - f(u_1')}{2}\right)}{\sinh \left(\frac{f(u_2) - f(u_1)}{2}\right) \sinh \left(\frac{f(u_2') - f(u_1')}{2}\right)}\,.\nonumber
\end{equation}
In quantum gravity we path integrate over this correlator using the Schwarzian action. This formula returns (see section \ref{sect:ads}).
%%%%%%%%%%%%%%%%%%%%%%%%%%%%%%%
\subsection{Bulk operator reconstruction}\label{sect:bulkrec}
This conceptual framework to compute bulk correlators in dS$_2$ has an intuitive interpretation in terms of bulk operator reconstruction \cite{Hamilton:2005ju,Hamilton:2006az,Kabat:2011rz,Kabat:2017mun}. 

Bulk reconstruction in dS on the gravitational saddle was studied in \cite{xiao2014holographic}. Working with continuous series representations \eqref{range} we write for dS$_2$
\begin{align}
    \nn\average{\varphi(u_1,u_2)\dots}&=C_\Delta\,\int_{u_1}^{u_2} d u\,K_\Delta (u\,\rvert\, u_1,u_2)\,\average{\mo_\Delta(u)\dots }\\&\qquad\qquad\qquad\qquad+C_{1-\Delta}\,\int_{u_1}^{u_2} d u\,K_{1-\Delta} (u\,\rvert\, u_1,u_2)\,\average{\mo_{1-\Delta}(u)\dots}.
\end{align}
Here the bulk to boundary propagator is 
\begin{align}
    K_\Delta(u\,\rvert\, u_1,u_2)=\frac{\sin^{\Delta-1}\left(\frac{u_2-u}{2}\right)\sin^{\Delta-1}\left(\frac{u-u_1}{2}\right)}{\sin^{\Delta-1}\left(\frac{u_2-u_1}{2}\right)}\,.
\end{align}
Note that we get contributions from two terms for continuous series representations, unlike in AdS \cite{Harlow:2011ke}. The prefactors in the expansion can be deduced by expanding the correlator near future infinity \cite{xiao2014holographic}. Given these prefactors, one confirms that the whole integral indeed recovers the whole correlator, via direct integration.\footnote{Initially one obtains the sum of two hypergeometric functions, which recombines correctly by using \cite{xiao2014holographic}
\begin{align}
    \nn \,_2F_1(\Delta,1-\Delta,1,z)&=(-1)^\Delta \frac{\Gamma(1-2\Delta)}{\Gamma(1-\Delta)^2}\,\frac{1}{z^\Delta}\,_2F_1(\Delta,\Delta,2\Delta,1/z)\\&\qquad + (-1)^{1-\Delta}\frac{\Gamma(2\Delta-1)}{\Gamma(\Delta)^2}\,\frac{1}{z^{1-\Delta}}\,_2F_1(1-\Delta,1-\Delta,2-2\Delta,1/z)\,.\label{identity}
\end{align}}

Away from the gravitational saddle, one can do bulk operator reconstruction in each of the backgrounds $g_f$ \cite{blommaert2019clocks}. The bulk to boundary propagator becomes
\begin{align}
    K_\Delta(u\,\rvert\, u_1,u_2)_f=\frac{1}{f(u)^{\Delta-1}}\,\frac{\sin^{\Delta-1}\left(\frac{f(u_2)-f(u)}{2}\right)\sin^{\Delta-1}\left(\frac{f(u)-f(u_1)}{2}\right)}{\sin^{\Delta-1}\left(\frac{f(u_2)-f(u_1)}{2}\right)}.\label{bubo}
\end{align}
The boundary correlator in the individual backgrounds is the integrand in \eqref{E:bdybdy1}, such that bulk operator reconstruction in any given background $g_f$ yields
\begin{align}
    \nn\average{\varphi(u_1,u_2)\dots}_f&=C_\Delta\,\int_{u_1}^{u_2} d u\,K_\Delta (u\,\rvert\, u_1,u_2)_f\,\average{\mo_\Delta(u)\dots }_f\\&\qquad\qquad\qquad\qquad+C_{1-\Delta}\,\int_{u_1}^{u_2} d u\,K_{1-\Delta} (u\,\rvert\, u_1,u_2)_f\,\average{\mo_{1-\Delta}(u)\dots}_f.\label{integ}
\end{align}
This reproduces the reparameterized correlator which is integrated over in \eqref{Wightman1}. Notably, the integration domain does not depend on the reparameterization, so we can compute the Schwarzian path integral of the integrand first. This enables an exact analysis, but does not simplify a perturbative one (see section \ref{sec:discussion} and \cite{blommaert2019clocks}).

The Schwarzian path integral of these products of boundary correlators in the integrand of \eqref{integ} should be interpreted as genuine transition amplitudes \eqref{wave}, with the final matter state determined by the locations of the fields on the future boundary, this follows from \cite{cotler2019emergent}. Therefore these bulk observables indeed decompose into metaobservables as claimed around \eqref{wave}.
%%%%%%%%%%%
% SECTION %
%%%%%%%%%%%%
\section{Encoding gravitational ordering}
\label{sec:shockandscramble}
The question arises how to compute out of time ordered bulk correlators in this framework. Suppose we consider two distinct massive scalars, in each fixed background $g_f$ the four point function obviously factorizes
\begin{align}
    \average{\varphi_1(u_1,v_1)\varphi_2(u_2,v_2)\varphi_1(u_3,v_3)\varphi_2(u_4,v_4)}_f&=\average{\varphi_1(u_1,v_1)\varphi_1(u_3,v_3)}_f\average{\varphi_2(u_2,v_2)\varphi_2(u_4,v_4)}_f\nn\,.
\end{align}
The relative ordering of $\varphi_2(u_2,v_2)$ and $\varphi_1(u_3,v_3)$ has zero influence in this matter calculation. Then how does the complete gravitational theory know about this relative ordering, why is
\begin{equation}
    \frac{1}{Z_\text{gravity}}\int[\dpi f]\,e^{i S_\text{gravity}[f]}\average{\varphi_1(u_1,v_1)\varphi_2(u_2,v_2)\varphi_1(u_3,v_3)\varphi_2(u_4,v_4)}_f\label{31}
\end{equation}
different from
\begin{equation}
     \frac{1}{Z_\text{gravity}}\int[\dpi f]\,e^{i S_\text{gravity}[f]}\average{\varphi_1(u_1,v_1)\varphi_1(u_3,v_3)\varphi_2(u_2,v_2)\varphi_2(u_4,v_4)}_f\,,
\end{equation}
with notably the same eight boundary points specifying the locations of the bulk operators?

In both computations we would write up the same classical matter correlation function as function of $g_f$. However in quantum gravity these classical functions of $g_f$ become operators. Concequently we must specify an operator ordering. These computations feature essentially two inequivalent quantum mechanical operators which coincide in the classical limit, like $\hat{p}\hat{x}$ and $\hat{x}\hat{p}$ but with gravitational operators.

The question is then what is the correct gravitational operator ordering that implements out of time ordered correlators in the bulk, and how do we implement it in practice? Here we dissect the computation of the bulk four point function in Schwarzian perturbation theory, and argue for such a prescription to compute out of time ordered correlators. Implementing this procedure in practice, we find maximal Lyapunov growth, as physically expected \cite{sekino2008fast,susskind2011addendum}.
%%%%%%%%%%%%%%%%%%%%%%%%%
\subsection{Proposal}
We propose to compute out of time ordered four point functions as follows. Suppose that all the four operators are inserted at the same angular coordinate and at times $t_1>t_2>t_3>t_4$. We compute
\begin{equation}
    \average{\varphi_1(t_1)\varphi_2(t_2)\varphi_1(t_3)\varphi_2(t_4)}=\quad
    \raisebox{-20mm}{\includegraphics[width=39mm]{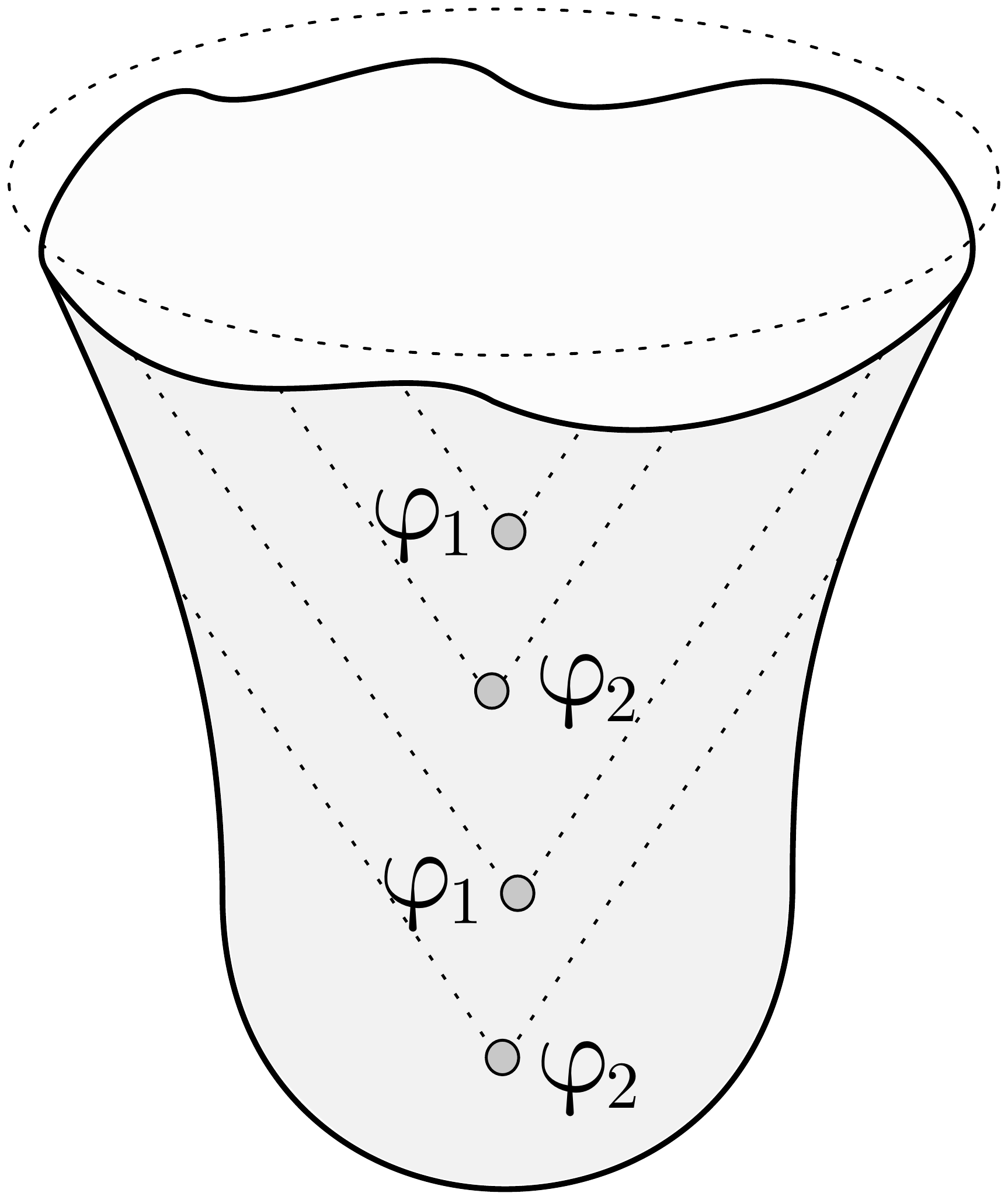}}\label{wrong}
\end{equation}
This represents an ordinary time ordered correlator, so we expect no surprises in computing it. We will learn that the perturbative Schwarzian correlator is determined by specifying the ordering of the eight Euclidean boundary coordinates which specify the bulk coordinates of the four involved bulk operators.

We propose that analytic continuation of the answer of this correlator to $t_1>t_3>t_2>t_4$ gives an out of time ordered correlator. In terms of the boundary coordinates, this means we analytically continue the correlator for given values of the eight points, to different Euclidean values of those eight points. 

We emphasize that while we will end up computing some Schwarzian eight point function, bulk reconstruction is not invoked, instead we directly compute bulk path integrals like \eqref{31}, with the analytic continuation motivated entirely from these bulk configurations.

The answer turns our very different to what one obtains by directly computing the time ordered correlator, given some identical set of coordinates $t_1>t_3>t_2>t_4$
\begin{equation}
    \average{\varphi_1(t_1)\varphi_1(t_3)\varphi_2(t_2)\phi_2(t_4)}=\quad
    \raisebox{-20mm}{\includegraphics[width=39mm]{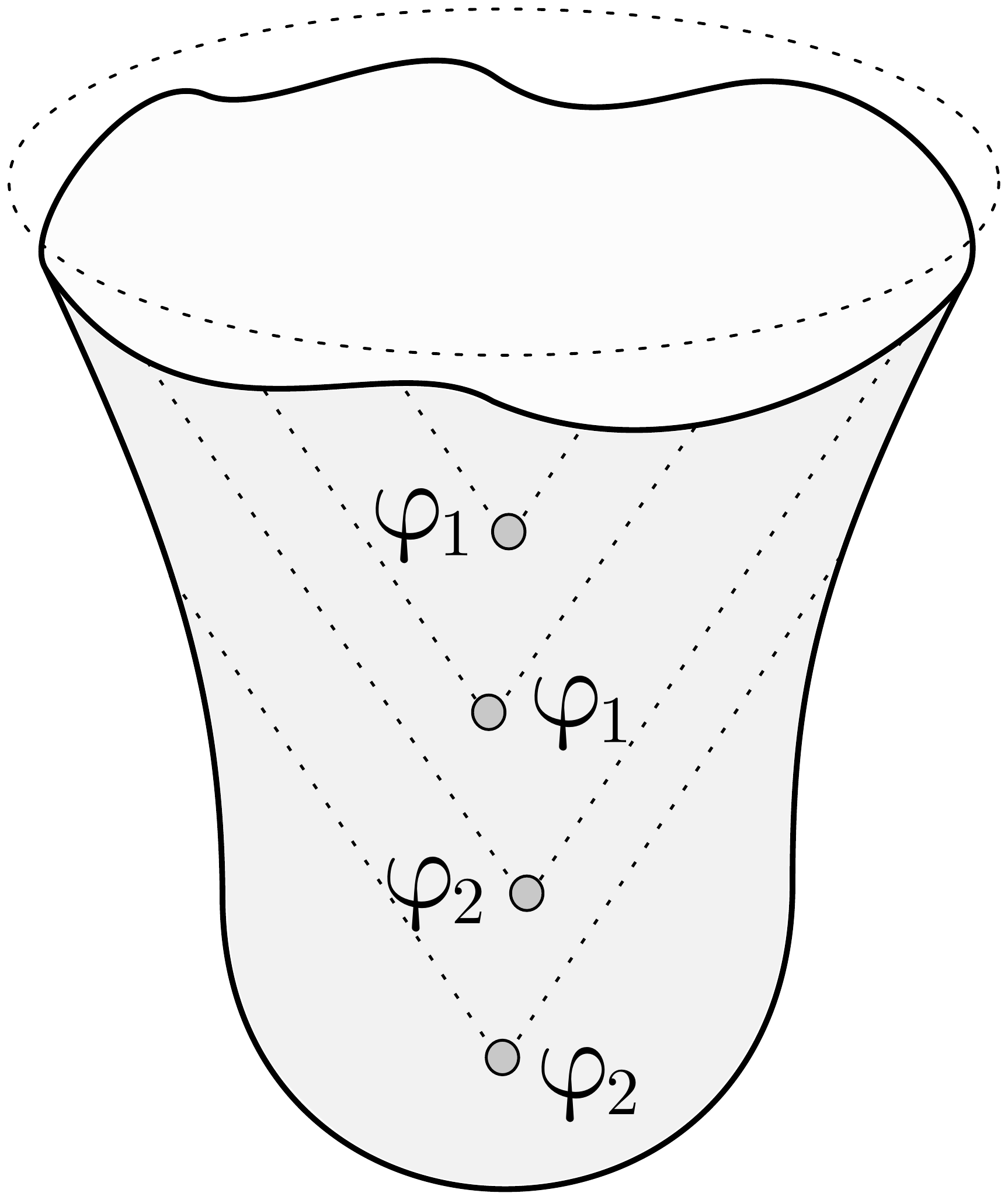}}\label{timeordered}
\end{equation}
The first grows exponentially in time at the maximal rate \cite{maldacena2016bound}, whilst the latter decays.

Since the time ordered and out of time ordered four point functions are specified by the same eight boundary points, the fact that we get a different answer implies that Euclidean perturbative Schwarzian correlators are multivalued functions, with branchcuts and different sheets. This is not obvious from the known formulas for perturbative Schwarzian correlators \cite{Maldacena:2016upp,Sarosi:2017ykf}, where it appears that perturbative Schwarzian correlators are just piecewise analytic functions. However we will show that each such piecewise analytic function should be viewed as the principle sheet of one function, with branchcut discontinuities and inequivalent sheets. Functions which are analytic in some open subset of the complex plain have unique analytic continuations, hence the piecewise analytic functions of \cite{Maldacena:2016upp,Sarosi:2017ykf} (for complex times) uniquely define the values of the Schwarzian correlators on these other sheets.

The branchcut discontinuities imply that the principle sheet of bulk correlators is divided into locally analytic patches, with branchcuts in between patches. These patches correspond with causally distinct locations of the bulk operators, for example \eqref{wrong} and \eqref{timeordered} correspond to inequivalent patches. 

We propose that analytic continuation from one such patch to another, thereby smoothly crossing branchcuts onto different sheets, implements other gravitational operator orderings. For general configurations the situation gets more complicated because we furthermore must track what happens with the branchcut in the hypergeometric matter propagator. However, when restricting to analytic continuation of \eqref{wrong} to $t_1>t_3>t_2>t_4$ we are isolating purely the gravitational analytic structure, since we are only affecting the relative ordering of $\varphi_2(t_2)$ and $\varphi_1(t_3)$. Evidently any chaotic dynamics is only due to gravitational backreaction effects, and not due to some structure in matter correlators on fixed backgrounds, therefore isolating effects associated with the gravitational analytic structure seems advisable.

In the remainder of this section we investigate this proposal. We write out this four point function in Schwarzian perturbation theory, and investigate its analytic structure, meaning branchcuts and sheets.
%%%%%%%%%%%%%%%%%%%%%%%%%%%%%%%%
\subsection{Perturbative expansion}
We want to compute the following out of time ordered correlator
\begin{equation}
    F(t)=\frac{\average{\varphi_1(t)\varphi_2(0)\varphi_1(t)\varphi_2(0)}}{\average{\varphi_2(0)\varphi_2(0)}\average{\varphi_1(t)\varphi_1(t)}}\,,\label{fone}
\end{equation}
with implicit regularizations to avoid probing on the singularity. This is the quantum gravity implementation of an out of time ordered correlator computed by some static patch observer.

The static patch is here (dependence on the temperature is suppressed, but relevant \cite{Aalsma:2020aib})
\begin{align}
    ds^2=-(1-r^2)\,dt^2 + \frac{d r^2}{(1-r^2)}\quad,\quad -1<r<1\,.\label{penrose3}
\end{align}
The static observer is interesting in this context, because he is quite similar to the asymptotic observer in AdS$_2$. When he releases some quanta, it accelerates towards the dS$_2$ horizon. The energy of the particle, as perceived by the static observer, grows exponentially as $\exp(2\pi t/\beta)$. Therefore we expect to find similar maximal Lyapunov growth in \eqref{fone}, following \cite{sekino2008fast,susskind2011addendum}.

We are interested in inserting operators at points $(t,r)$, but we have seen this is no way of specifying bulk points in quantum gravity. Bulk points are to be defined using boundary points, and geometric constructions. On the gravitational saddle we can map $(t,r)$ to two boundary points $(u_1,u_2)$, by determining where light sent from our point reaches the future boundary. Specifying two boundary points $(u_1,u_2)$ accompanied by this lightray prescription does make sense as definition of a bulk point in quantum gravity, so we will use this definition for the point $(t,r)$ in what follows.

The standard out of time ordered correlator \eqref{fone} is inserting operators in the center of the static patch. We emphasize that this is not identical to the line $\theta=0$, but instead is the line $u_1+u_2=0$ in \eqref{210}. The later is a diffeomorphism invariant statement. This might be slightly confusing in this two dimensional setup, since all metrics are reparameterizations of one another, but in more complicated setups it should be clear that $\theta=0$ has no meaning in quantum gravity. This means the analysis of \cite{Aalsma:2020aib} on out of time ordered correlators in AdS$_3$, is somewhat pathological. Inserting operators at fixed classical bulk locations is computing some diff covariant quantity in quantum gravity, and is therefore nonphysical. 

Using \eqref{Wightman1} we compute the nominator of \eqref{fone} as
\begin{align}
    &\average{\varphi_1(v_1,u_1)\varphi_2(v_2,u_2)\varphi_1(v_3,u_3)\varphi_2(v_4,u_4)}\nn\\&\qquad \qquad =\frac{1}{Z_\text{gravity}}\int [\dpi f]\,e^{-S_\text{gravity}[f]} \,_2F_1(\Delta,1-\Delta,1,z^{13}_f))\,_2F_1(\Delta,1-\Delta,1,z^{24}_f)  \,.\label{335}
\end{align}
For the moment let us consider generic bulk points, we have for example
\begin{equation}
    z^{24}_f=\frac{\sin\left( \frac{f(u_2)-f(v_4)}{2}\right)\sin\left( \frac{f(u_4)-f(v_2)}{2}\right)}{\sin\left( \frac{f(u_2)-f(v_2)}{2}\right)\sin\left( \frac{f(u_4)-f(v_4)}{2}\right)}\,.\label{337}
\end{equation}
We expect to find Lyapunov growth at linear order in the effective Newton constant $G\ell$
\begin{equation}
    1-F(t)\sim G\ell\, \exp(2\pi t/\beta)\,.\label{329}
\end{equation}
This leads us to calculate the order $G\ell$ contribution to \eqref{335} using Schwarzian perturbation theory. 

We can expand $f(u)$ near the classical saddle as
\begin{equation}
    f(u)=u + \varepsilon(u)\,.
\end{equation}
Expanding the Schwarzian action \eqref{E:action2} one finds to leading order in $\varepsilon(u)$ a quadratic action, with a prefactor proportional to $1/G\ell$ \cite{Sarosi:2017ykf,Saad:2019lba}. The zeroth order term in the Taylor expansion in $\varepsilon(u)$ of the inserted matter correlators gives the classical answer. Linear terms in $\varepsilon(u)$ do not contribute, and quadratic terms gives a contribution proportional to $G\ell$. This expansion in $G\ell$ breaks down when the corrections become order one, in our case this happens when
\begin{equation}
    t\sim \frac{\beta}{2\pi}\,\ln \frac{1}{G\ell}=t_*\,,\label{332}
\end{equation}
the scrambling time in this context. 

To confirm or correct \eqref{329}, we just expand \eqref{335} to quadratic order in $\varepsilon(u)$ and compute the Gaussian path integrals. Imagine Taylor expanding the product of the two hypergeometric functions in \eqref{335} to quadratic order in $\varepsilon(u)$. The quadratic term gets two contributions. The first is the sum of the quadratic terms in the individual expansion of each of the hypergeometric functions, and the second is the product of the linear pieces in those expansions. Only the latter connected contribution ends up contributing at order $G\ell$ in \eqref{fone}. The first contribution cancels with identical terms from the denominator of \eqref{fone}.

It is useful for future purposes to organize the resulting calculation as follows. We write the perturbative expansion of $z_f^{13}$ to linear order in $\varepsilon(u)$ as
\begin{equation}
    \frac{z_f^{13}}{z^{13}}=1+\frac{1}{2}\,L(u_1,u_3,v_1,v_3)\,.
\end{equation}
This linear function in $\varepsilon(u)$ is a sum of four simple terms
\begin{equation}
    L(u_1,u_3,v_1,v_3)=L(u_1,v_1)+L(u_3,v_3)-L(u_1,v_3)-L(u_3,v_1)\,,\label{signs}
\end{equation}
with each of the simple terms given by
\begin{equation}
    L(u_1,u_2)=\varepsilon'(u_1)+\varepsilon'(u_2)-\frac{\varepsilon(u_1)-\varepsilon(u_2)}{\tan \left(\frac{u_1-u_2}{2}\right)}\,.\label{pertL}
\end{equation}
Notice that this is the same as the linear correction to a Schwarzian boundary bilocal \cite{Maldacena:2016upp}
\begin{equation}
   \frac{f'(u) f'(v)}{\sin^{2} \left(\frac{f(u)-f(v)}{2}\right)}=\frac{1}{\sin^{2} \left(\frac{u-v}{2}\right)}+ L(u,v)\,\frac{1}{\sin^{2} \left(\frac{u-v}{2}\right)}\,.\label{perttwo}
\end{equation}
To linear order in the Schwarzian perturbation the hypergeometric propagator is then
\begin{align}
    G(z^{13}_f)=G(z^{13})+\frac{z^{13}}{2}G'(z^{13})\,L(u_1,u_3,v_1,v_3)\,.
\end{align}
Combining the elements, we find for the out of time ordered correlator
\begin{equation}
    F(t)=1+\frac{z^{13}z^{24}}{4}\frac{G'(z^{13})}{G(z^{13})}\frac{G'(z^{24})}{G(z^{24})}\average{L(u_1,u_3,v_1,v_3)L(u_2,u_4,v_2,v_4)}\,,\label{pert}
\end{equation}
where the last factor is Gaussian path integral over $\varepsilon(u)$. In terms of operator ordering, we can focus all of our attention on this last factor, which decomposes into 16 terms of the type
\begin{equation}
\average{L(x_1,x_2)L(x_3,x_4)}\,. \label{four}
\end{equation}
Because of \eqref{perttwo}, these factors are precisely identical to the perturbative Schwarzian four point functions computed in the context of AdS$_2$ in \cite{Sarosi:2017ykf,Maldacena:2016upp,Cotler:2019nbi}. This is convenient, it means we can essentially ship in their formulas and investigate their analytic structure.

For future purposes we repeat this analysis for bulk (out of time ordered) correlators in AdS$_2$. We are primarily interested in cases where the operators $\varphi_1(v_2,u_2)$ and $\varphi_1(v_4,u_4)$ are inserted close together together. The relevant propagator can be rewritten as \cite{blommaert2019clocks}
\begin{equation}
    \tilde{G}(w)\sim \frac{1}{\left(1-2w\right)^{\ell}} \,_2F_1\left(\frac{\ell}{2},\frac{\ell+1}{2},\frac{2\ell+1}{2},\frac{1}{\left(1-2w\right)^{2}}\right)\,,
\end{equation}
modulo irrelevant prefactors. Here the crossratio is
\begin{equation}
    w_f^{24}=\frac{\sin\left(\frac{f(u_2)-f(u_4)}{2}\right)\sin\left(\frac{f(v_2)-f(v_4)}{2}\right)}{\sin\left(\frac{f(u_2)-f(v_2)}{2}\right)\sin\left(\frac{f(u_4)-f(v_4)}{2}\right)}.
\end{equation}
We also Wick rotated from Lorentzian to Euclidean boundary coordinates. To linear order in the Schwarzian perturbation we have
\begin{equation}
    \frac{w_f^{24}}{w_f^{24}}=1+\frac{1}{2}\tilde{L}(u_2,u_4,v_2,v_4)\,,
\end{equation}
where now
\begin{equation}
    \tilde{L}(u_2,u_4,v_2,v_4)=L(u_2,v_2)+L(u_4,v_4)-L(u_2,u_4)-L(v_2,v_4)\,.
\end{equation}
The bulk out of time ordered correlators in AdS$_2$ are computed to leading order in $G\beta$ as
\begin{equation}
    F(t)=1+\frac{w^{13}w^{24}}{4}\frac{\tilde{G}'(w^{13})}{\tilde{G}(w^{13})}\frac{\tilde{G}'(w^{24})}{\tilde{G}(w^{24})}\langle\tilde{L}(u_1,u_3,v_1,v_3)\tilde{L}(u_2,u_4,v_2,v_4)\rangle\,.\label{adspert}
\end{equation}
%%%%%%%%%%%%%%%%%%%%%%%%%%%%%%%%%
\subsection{Schwarzian sheets}
We now show that perturbative Euclidean Schwarzian correlators are multivalued functions with branchcuts, and explain our prescription for computing out of time ordered correlators. Consider \eqref{wrong} and \eqref{timeordered}, with perturbative expansion \eqref{pert}, what is the analytic structure of this perturbative expansion depending on the relative ordering of $t_2$ and $t_3$?

Remember that this setup does not probe the analytic structure of the hypergeometric prefactors in \eqref{pert}, by construction, so any interesting analytic structure, for our purposes, should be entirely encoded in the perturbative Schwarzian eight point correlator
\begin{equation}
    \average{L(u_1,u_3,v_1,v_3)L(u_2,u_4,v_2,v_4)}\,.\label{ll}
\end{equation}
This correlator vanishes when gravitational interactions are turned off, hence we are probing pure gravitational backreaction effects by investigating this analytic structure, as we should.

This correlator decomposes into a linear combination of correlators of the type \eqref{four}, which are perturbative Schwarzian four point functions, hence we can focus on their analytic structure. Via \eqref{pertL} these decompose into linear combinations of (derivatives of) correlators like $\average{\varepsilon(u_1)\varepsilon(u_2)}$. The tangent factors in \eqref{pertL} can be ignored in terms of analytic structure, these have branches nor sheets. Also there is no pole when $t_2$ approaches $t_3$ in \eqref{wrong} or \eqref{timeordered}.

Hence any interesting structure, for our purposes, is in correlators of the type $\average{\varepsilon(u_1)\varepsilon(u_2)}$. Thanks to translation symmetry we can focus on $\average{\varepsilon(u)\varepsilon(0)}$, which was computed in \cite{Maldacena:2016upp,Sarosi:2017ykf}. Parameterizing $u=x+i y$, the complex annulus is covered by $-\pi<x<\pi$, and the answer for this elementary correlator can be written as \cite{Maldacena:2016upp,Sarosi:2017ykf}
\begin{equation}
    \frac{1}{2 G\ell}\,\average{\varepsilon(u)\varepsilon(0)}=2+\frac{\pi^2}{3}+5\cos(u)-(u-\text{sgn}(x)\pi)^2+2\sin(u)\,(u-\text{sgn}(x)\pi)\,,\label{eps}
\end{equation}
which is smooth on the whole complex annulus except on the imaginary axis, where we have a discontinuity. Because of this jump, the time ordered and out of time ordered perturbative Schwarzian four point functions, distinguished by the ordering of $x_1,x_2,x_3$ and $x_4$ in \eqref{four} along the Euclidean boundary, are different analytic functions \cite{Maldacena:2016upp}.

These can be recognizes as different sheets of one function with branchcuts, by rewriting
\begin{equation}
    \frac{1}{2 G\ell}\,\average{\varepsilon(u)\varepsilon(0)}=2+\frac{\pi^2}{3}+5\cos(u)+\ln(-\exp(i u))^2-2\, i \sin(u)\,\ln(-\exp(i u))\,.\label{eps2}
\end{equation}
Plotting the imaginary part of $\ln(-\exp(i u))$ gives a helix structure (identify the boundaries)
\begin{equation}
    \quad
    \raisebox{-10mm}{\includegraphics[width=48mm]{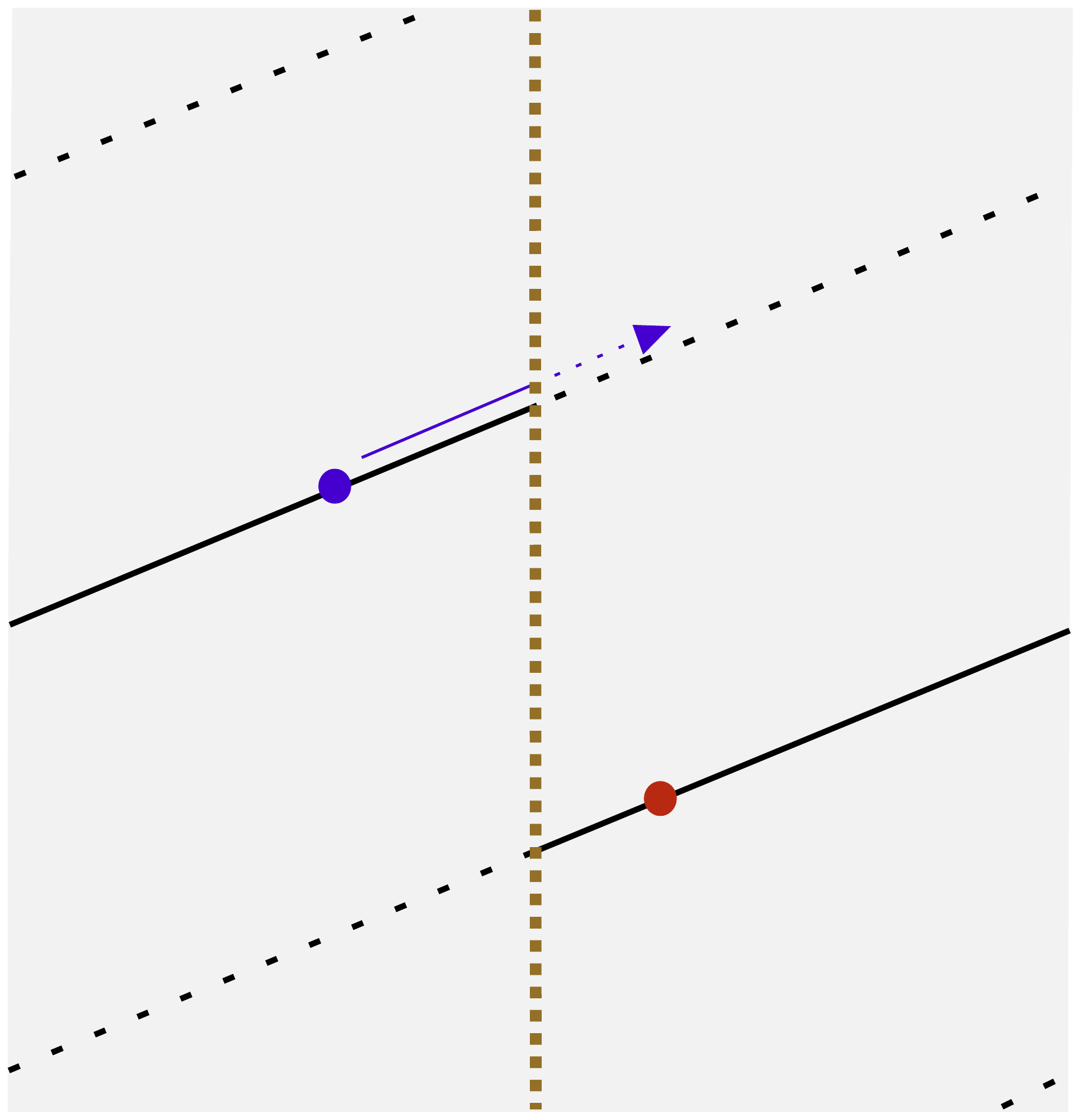}}\quad
\end{equation}
This is independent of the real part $y$. Continuation from the principle sheet (black), through the branchcut (zipper) takes us onto other sheets (dotted black). There is an infinite number of sheets, but in practice we are only interested in two of them, we are interested in local manipulations and not in moving points all the way around the Euclidean boundary circle. 

The principle sheet is like some stack of the helix.

From this picture we see immediately that \eqref{eps2} takes the form \eqref{eps} on the principle sheet, and therefore it represents the unique analytic continuation of \eqref{eps} to other sheets.

This means the correlator $\average{\varepsilon(u_1)\varepsilon(u_2)}$ is effectively double valued with analytic structure on the complex annulus that can be graphically represented as
\begin{equation}
    \quad
    \raisebox{-10mm}{\includegraphics[width=42mm]{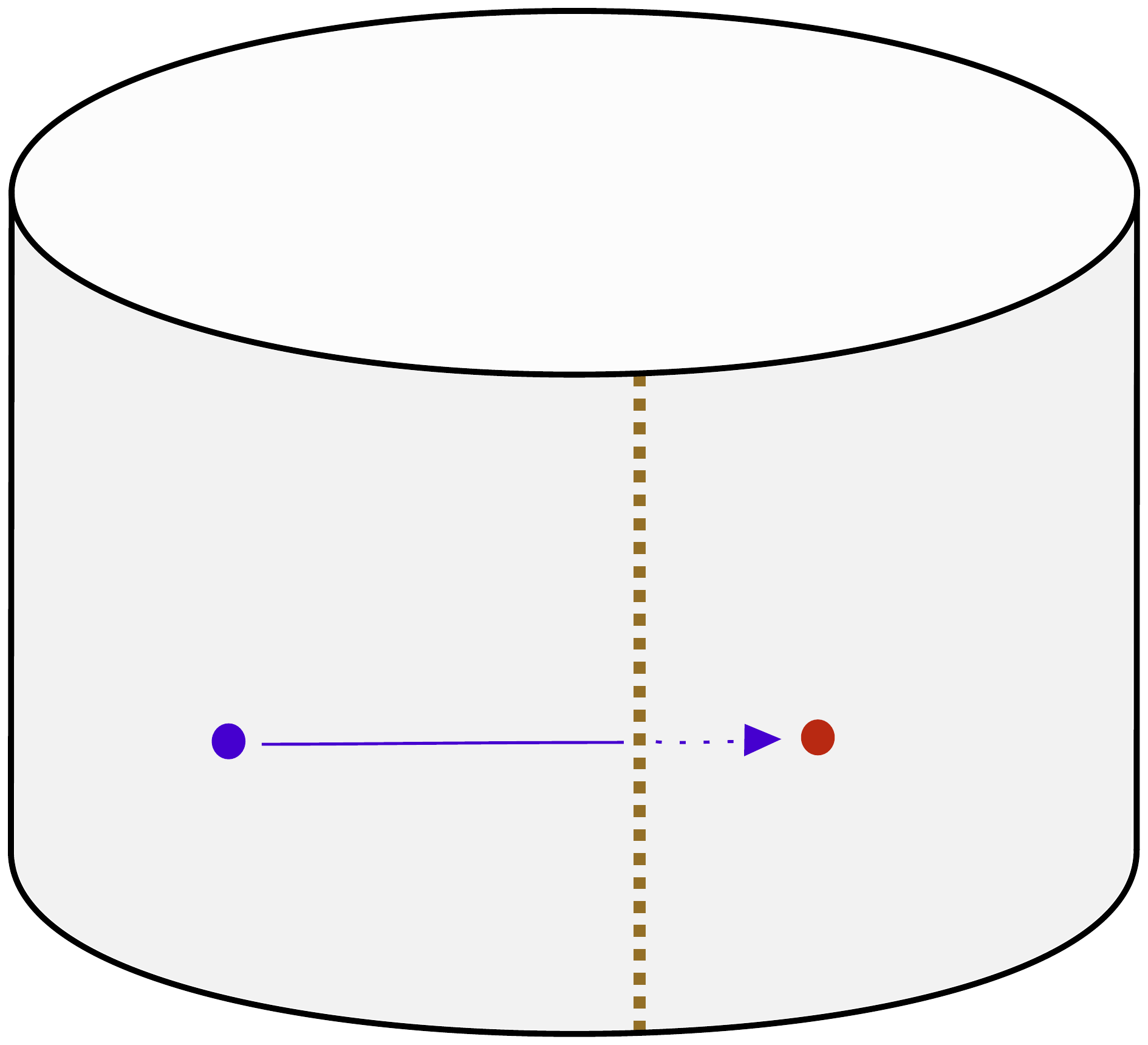}}\quad\label{struc}
\end{equation}
One obtains the correlator $\average{\varepsilon(x)\varepsilon(0)}$ for $x>0$ on the second sheet by analytic continuation through the branchcut starting from the patch $x<0$ (blue). More general, the second sheet of \eqref{eps2} is obtained through analytic continuation of \eqref{eps} through this branchcut
\begin{equation}
    \frac{1}{2 G\ell}\,\average{\varepsilon(u)\varepsilon(0)}=2+\frac{\pi^2}{3}+5\cos(u)-(u+\text{sgn}(x)\pi)^2+2\sin(u)\,(u+\text{sgn}(x)\pi)\,.\label{epsbis}
\end{equation}

The perturbative Schwarzian four point functions \eqref{four} compose of sums of (derivatives of) correlators like $\average{\varepsilon(u_1)\varepsilon(u_2)}$, hence they inherent this sheet structure and are multivalued too. Consider as an example the perturbative AdS$_2$ boundary four point function \cite{Maldacena:2016upp}, which is essentially \eqref{four}. Suppose $x_1>x_2>x_4$ are fixed and consider $x_3$ close to $x_2$. The analytic structure of the correlator is then locally identical to \eqref{struc} 
\begin{equation}
    \quad
    \raisebox{-10mm}{\includegraphics[width=42mm]{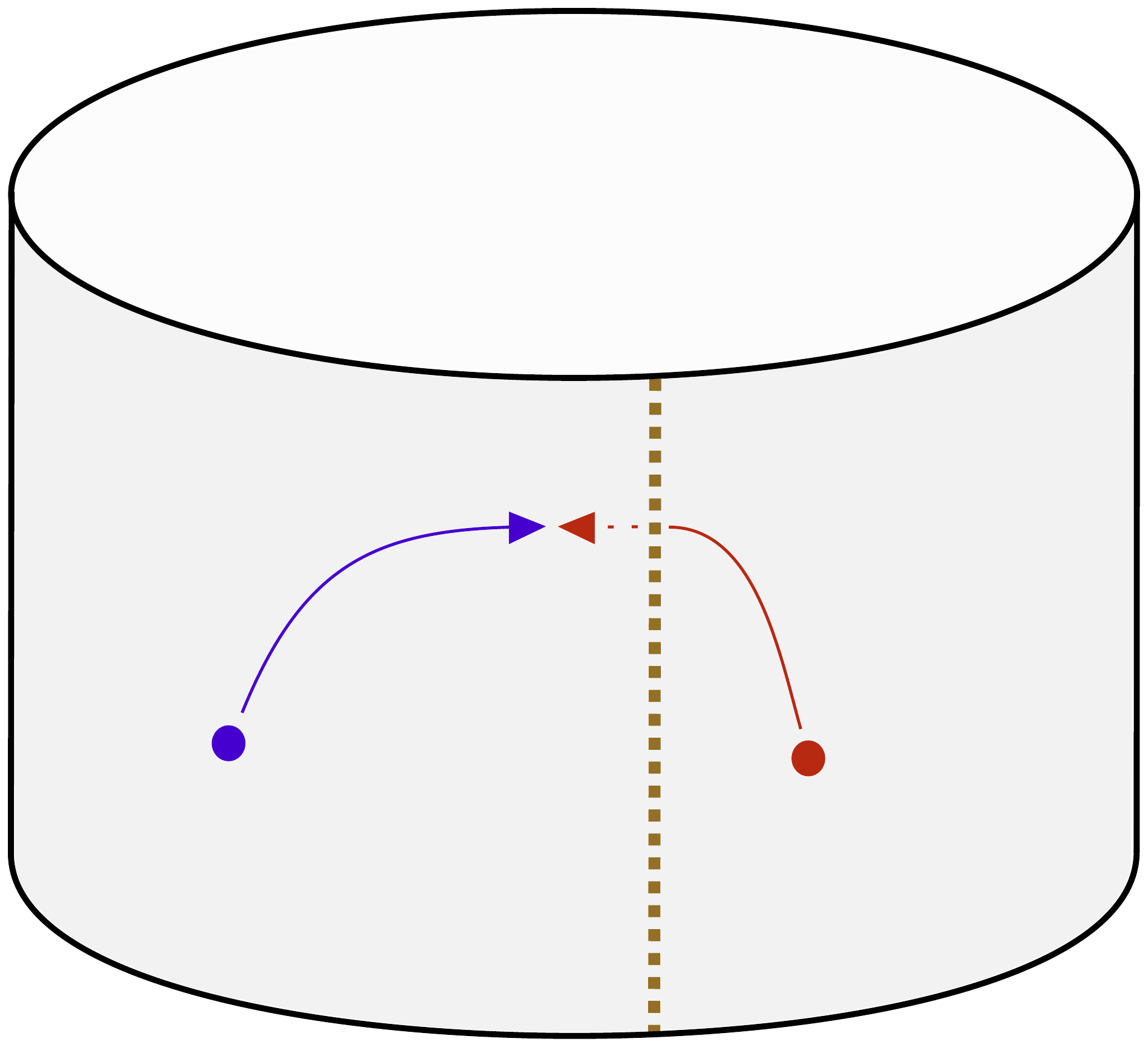}}\quad\label{adsanco}
\end{equation}
The time ordered correlator is found via analytic (red) continuation of the correlator on the principle sheet for $x_1>x_2>x_3>x_4$ to imaginary values, and analytic continuation from the region $x_1>x_3>x_2>x_4$ gives the out of time ordered correlator (blue). Typically, one computes both on the principle sheet by introducing a Euclidean regularization, because in quantum mechanics is is nonphysical to consider propagation for negative Euclidean times. However in this context, the Schwarzian boundary correlators are just expressions that arise in the evaluation of bulk observables, and therefore we are not constrained by requiring some natural interpretation of Schwarzian correlators in quantum mechanics, so that these sheets are fair game for our investigation of bulk correlators.

We now know how to practically implement the calculation of the dS$_2$ bulk out of time ordered four point function \eqref{wrong} with $t_1>t_3>t_2>t_4$. First one calculates
\begin{equation}
    \quad
    \raisebox{-20mm}{\includegraphics[width=39mm]{otoc39.pdf}}
\end{equation}
for $t_1>t_2>t_3>t_4$, which through \eqref{pert} reduces to computing the Schwarzian eight point function \eqref{ll} on the principle sheet with Euclidean coordinates ordered like
\begin{equation}
    u_4>u_3>u_2>u_1>v_1>v_2>v_3>v_4\,.\label{orderone}
\end{equation}
To get an out of time ordered correlator we analytically continue the answer of this Schwarzian eight point function to different values for these eight Euclidean coordinates, corresponding with $t_1>t_3>t_2>t_4$. This trajectory crosses several branchcuts thus this computes \eqref{ll} on some nonprinciple sheet, the value on the first sheet computes \eqref{timeordered} with $t_1>t_3>t_2>t_4$. Therefore the technical explanation why time ordered and out of time ordered bulk four point functions are different in dS$_2$, is because one evaluates the Schwarzian correlators like \eqref{ll} on different sheets, depending on the gravitational operator ordering.

Notice for clarity that the infinite extend of the branchcut in \eqref{eps2} along the imaginary axis, divides the principle sheet of \eqref{ll} into disconnected patches bounded by branchcuts, with different patches corresponding to causally inequivalent locations of the bulk operators.
%%%%%%%%%%%%%%%%%%%%%%%%%%%%%%%
\subsection{Maximal chaos}\label{sect:otoc1}
We remind the reader that we want to compute 
\begin{equation}
    F(t)=\frac{\average{\varphi_1(t)\varphi_2(0)\varphi_1(t)\varphi_2(0)}}{\average{\varphi_2(0)\varphi_2(0)}\average{\varphi_1(t)\varphi_1(t)}}\,,\label{336}
\end{equation}
which perturbatively becomes \eqref{pert}. This means we should calculate \eqref{ll}, with operators ordered as in \eqref{orderone}, which means summing up perturbative Schwarzian four point functions. We then analytically continue the answer to coordinates $u_2,u_3,v_2$ and $v_3$ as specified by the locations of the bulk operators in \eqref{336}. 

We introduce numerically small radial (in static coordinates) regulators to avoid identical operators on identical locations, and obtain eventually (see appendix \ref{app:fourpoint})
\begin{equation}
    \average{L(u_1,u_3,v_1,v_3)L(u_2,u_4,v_2,v_4)}\sim \varepsilon_1\,\varepsilon_2\,G\ell \exp(2\pi t/\beta)\quad,\quad \frac{2\pi}{\beta}\ll t\,.\label{340}
\end{equation}
We restored dependence on the de Sitter length via the temperature. We can evaluate \eqref{336} by inserting into \eqref{pert} the value of the propagator for nearby operators, for $m^2\gg 1$ this is
\begin{equation} 
    G(z^{13})\sim \ln \frac{4}{\varepsilon_2^2}.\label{Wightmanper}
\end{equation}
Therefore we arrive at
\begin{equation}
    1-F(t)\sim -\frac{1}{\varepsilon_1\ln \varepsilon_1}\,\frac{1}{\varepsilon_2\ln \varepsilon_2}\,G\ell \exp(2\pi t/\beta)\quad,\quad \frac{2\pi}{\beta}\ll t\ll \frac{\beta}{2\pi} \ln \frac{1}{G\ell}\,. \label{f11}
\end{equation}
This is our main conclusion, and provides evidence from the gravitational path integral that dS JT gravity is maximally chaotic within our choice of framework, and given this proposal for encoding out of time ordering in bulk correlators. 

This solidifies the expectation that cosmological horizons are maximally chaotic \cite{sekino2008fast,susskind2011addendum}. The sign here is not important, and depends on if we choose spacelike or timelike regulation, the essential physical content of these single sided out of time ordered correlators is to prove that the commutator squared grows exponentially with time as $\sim \exp(2\pi t/\beta)$, the numerical size and the sign of the prefactor are not universal, and therefore less important \cite{Sarosi:2017ykf,maldacena2016bound,Polchinski:2015cea}

These calculations generalizes in a straightforward way, one could consider generic bulk points. We limit ourselves to one other particularly natural example, the two sided correlator \cite{shenker2014black}. This is natural, because we can compute it experimentally by perturbing the thermofield double state at some time $t$ in the past, before measuring the expectation value of some two sided correlator in the perturbed state. Standard shockwave reasoning suggests this should grow exponentially as $\exp(2\pi t/\beta)$, interested readers are referred to \cite{Aalsma:2020aib,Geng:2020kxh} for details.

For this experiment we compute 
\begin{equation}
     F(t)=\frac{\average{\varphi_2^R(-t)\varphi_1^R(0)\varphi_1^L(0)\varphi_2^R(-t)}}{\average{\varphi_1^R(0)\varphi_1^L(0)}\average{\varphi_2^R(-t)\varphi_2^R(-t)}}\,.\label{ftwo0}
\end{equation}
Appropriate regulators are left implicit. We should still use \eqref{pert}, however we first calculate
\begin{equation}
    \quad
    \raisebox{-20mm}{\includegraphics[width=39mm]{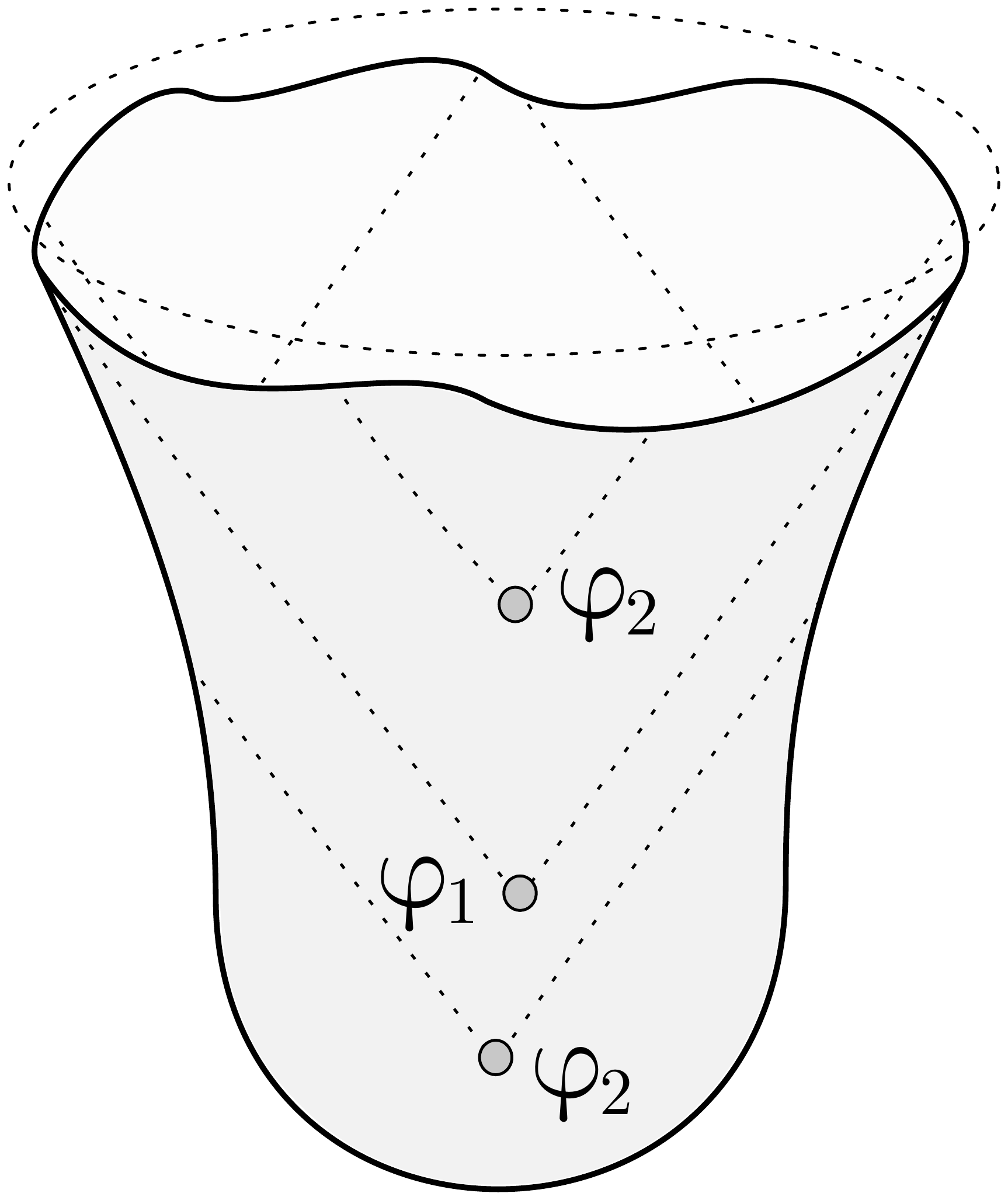}}\quad\text{(frontside)}\quad,\quad \raisebox{-20mm}{\includegraphics[width=39mm]{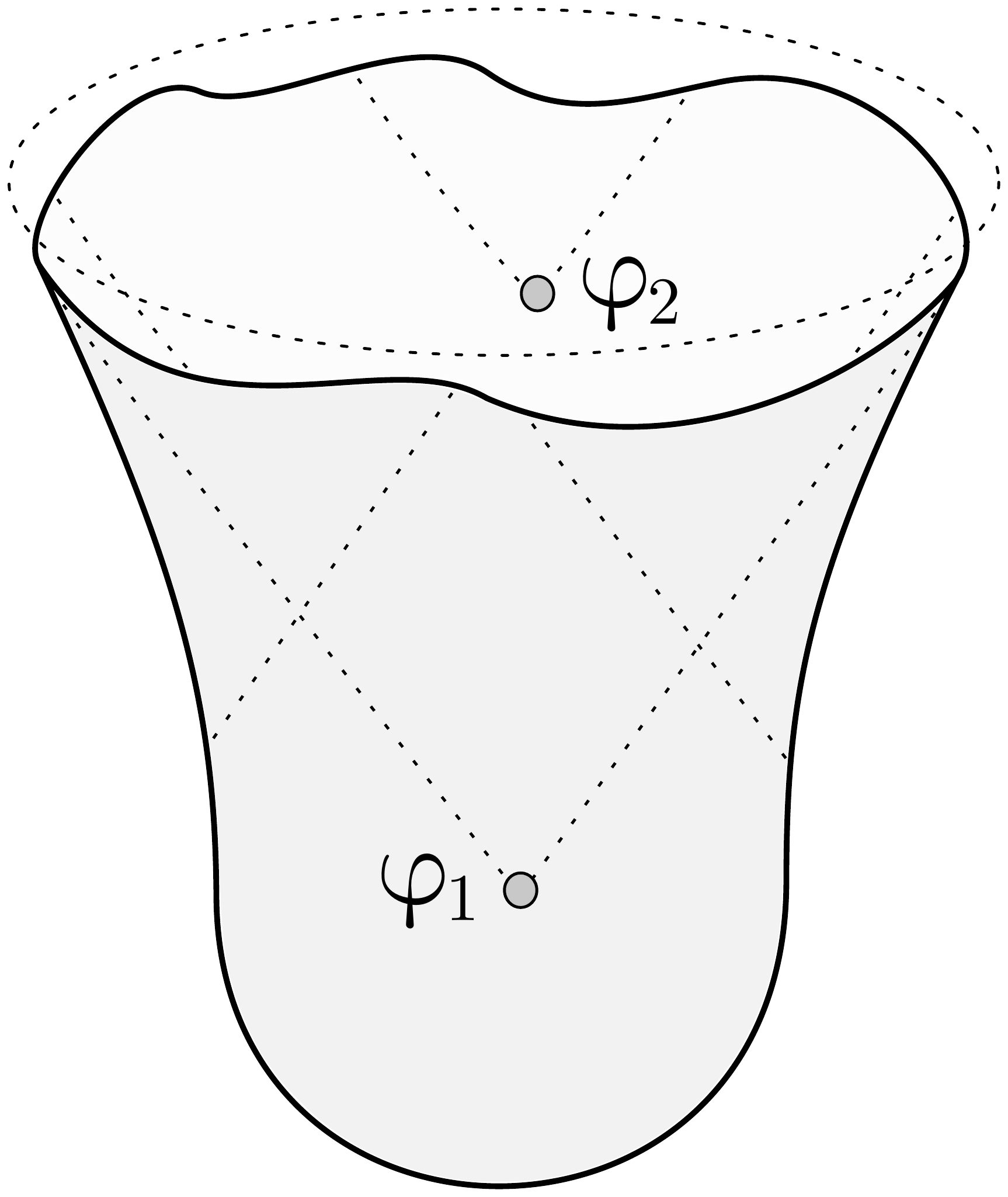}}\quad\text{(backside)}\quad
\end{equation}
The answer for \eqref{ll} with this ordering of the boundary points is then analytically continued by replacing $t_1>0$ with $-t$. After an elementary calculation, one obtains exponential growth
\begin{equation}
    1-F(t)\sim 
    \frac{\varepsilon_1}{\varepsilon_2 \ln \varepsilon_2}\,G\ell \exp(2\pi t/\beta) \quad,\quad \frac{2\pi}{\beta}\ll t\ll \frac{\beta}{2\pi} \ln \frac{1}{G\ell}\,. \label{ftwoans}
\end{equation}
This matches the answer from a shockwave calculation, including regulator dependence (see section \ref{sec:discussion}). Remember, however, that any semiclassical shockwave calculation in this context is gauge covariant, hence wrong. Gravitational dressing contribute at the same order so one should not bother comparing details, like prefactors.

Note the two sided correlation grows with time, as follows from shockwave intuition \cite{Aalsma:2020aib}. This reflects the fact that early perturbations create traversable wormholes in dS, unlike in AdS, where shockwaves generically make spatial wormholes elongate \cite{shenker2014black,shenker2014multiple}.
%%%%%%%%%%%%%%%%%%%%%%%%%%%%%
%%%%%%%%%%%
% SECTION %
%%%%%%%%%%%
\section{AdS butterflies}\label{sect:ads}
Using these same techniques, we can compute bulk out of time ordered correlators in AdS$_2$. Working with thermal coordinates \eqref{cooads}, our goal is to calculate for $t_2>t_1$ observables like
\begin{equation}
    F(t_1,z_1,t_2,z_2)=\frac{\average{\varphi_2(t_2,z_2)\varphi_1(t_1,z_1)\varphi_2(t_2,z_2)\varphi_1(t_1,z_1)}}{\average{\varphi_1(t_1,z_1)\varphi_1(t_1,z_1)}\average{\varphi_2(t_2,z_2)\varphi_2(t_2,z_2)}}\,,\label{41}
\end{equation}
which perturbatively becomes \eqref{adspert}, again regulators are left implicit. We focus on timelike separated operators, so that the (unfolded) Lorentzian time contour looks like the following
\begin{equation}
    \raisebox{-10mm}{\includegraphics[width=71mm]{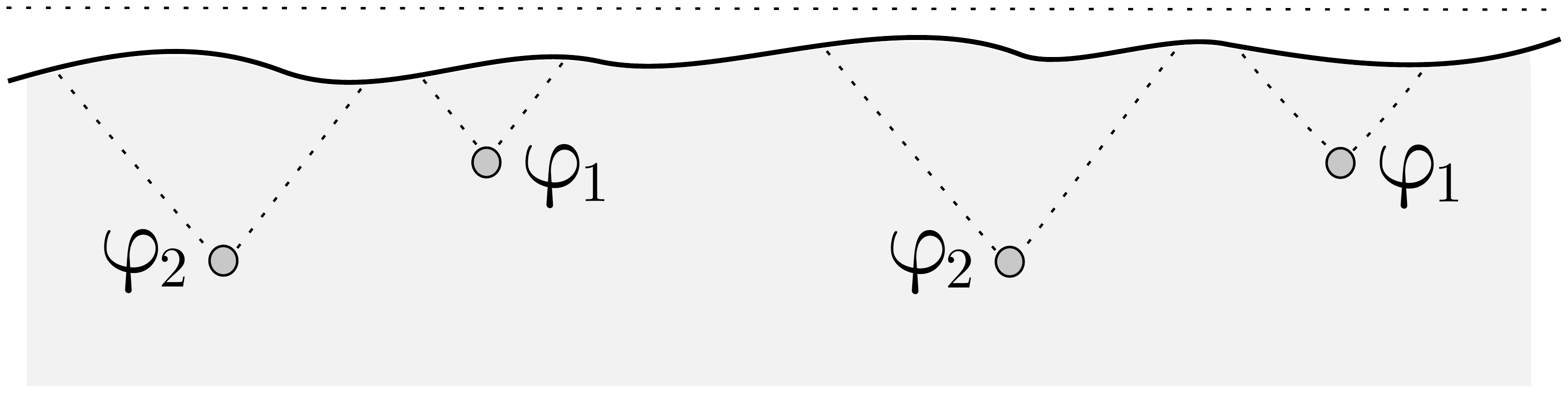}}
\end{equation}
The perturbative Schwarzian path integral that features in \eqref{adspert} is computed initially as a Euclidean correlator, which we then analytically continue to these eight Lorentzian boundary times.

We must specify how to order those Euclidean boundary points in the initial calculation. Intuition from the above picture suggests to consider
\begin{equation}
    u_1>v_1>u_2>v_2>u_3>v_3>u_4>v_4\,.
\end{equation}
We then simply analytically continue to the Lorentzian times, specified in \eqref{41}. Introducing small timelike regulators, an elementary calculation for large enough $z_1$ and $z_2$ results in the behavior
\begin{equation}
    \langle\tilde{L}(u_1,u_3,v_1,v_3)\tilde{L}(u_2,u_4,v_2,v_4)\rangle\sim \frac{1}{\delta_1\,\delta_2}\,G\beta\,\exp(2\pi (t_2-t_1+z_2+z_1)/\beta)\,.
\end{equation}
The only additional required ingredient is the propagator for nearby operators, in this case
\begin{equation}
G(w^{13})\sim \ln \frac{4\exp(2z_1)}{\delta_1^2}\,.\label{46real}    
\end{equation}
Therefore, one finally obtains (the regulators are just numerically small, not parameterically)
\begin{equation}
    1-F(t_1,z_1,t_2,z_2)\sim \frac{2\pi\beta}{z_1}\frac{2\pi\beta}{z_2}\frac{1}{\delta_1\,\delta_2}\,G\beta\,\exp(2\pi (t_2-t_1+z_2+z_1)/\beta)\,,\label{47real}
\end{equation}
which we can trust until the time difference approaches the scrambling time. The prefactor dependence on $z_1$ and $z_2$ is from \eqref{46real}. It is straightforward to repeat this for massless bulk fields using formulas of \cite{blommaert2019clocks}, leading to identical exponential growth.

This answer matches two pieces of intuition, which support this framework \cite{blommaert2019clocks}.

Bulk operator reconstruction writes bulk operators as linear combinations of boundary operators, therefore this bulk calculation can be viewed as a linear combination of ordinary boundary out of time ordered correlators, each growing exponentially with time. The leading contribution comes from the boundary operators separated by the longest times. Notice that the exponent in \eqref{47real} is exactly this maximal time difference corresponding to $\varphi_1(t_1,z_1)$ and $\varphi_2(t_2,z_2)$ bulk operator decomposition, as consistency demands.

Shockwave logic dictates the general form of this out of time ordered correlator should be $1-F(t_1,t_2,z_1,z_2)\sim E_1(t_1,z_1,t_2)E_2(z_2)$, where $E_1(t_1,z_1,t_2)$ is the energy of some shockwave created at $(t_1,z_1)$, as measured by some observer at $(t_2,z_2)$, and $E_2(z_2)$ is the local energy of the particle used to detect said shockwave. By rewriting this in terms of asymptotic energies, we reproduce the above answer $1-F(t_1,z_1,t_2,z_2)\sim E_1(0)E_2(0)\,\exp(2\pi(t_2-t_1+z_2+z_1)/\beta)$.
%%%%%%%%%%%
% SECTION %
%%%%%%%%%%%
\section{Conclusion}
\label{sec:discussion}
We presented and worked through a proposal for implementing out of time ordered correlators in dS JT gravity, and found exponential growth $\sim \exp(2\pi t/\beta)$. This is the behavior we expect for an out of time ordered correlator, which is an a posteriori motivation for this proposal. Recovering exponential growth $\sim \exp(2\pi t/\beta)$ should be viewed as a constraint on proposals for computing out of time ordered correlators in quantum gravity.

It would be interesting to try to compute bulk out of time ordered correlators in another framework for observables in cosmology \cite{Maldacena:2019cbz}, the analytic continuation we explained around \eqref{wrong} should work, but calculating bulk four point functions in their framework could prove more technically challenging.

It is not clear whether $\lambda=2\pi/\beta$ is an upper bound on chaos in this cosmological context, so it would be interesting to compute corrections to the Lyapunov exponents of cosmological horizons and to check whether these satisfy or violate the bound of \cite{maldacena2016bound}.

One could wonder what this analytic continuation of bulk correlators implies in terms of bulk operator reconstruction. It seems likely that the dual description involves propagation for negative Euclidean distances, which sounds quirky, but the dual of dS may well be quirky.
\\~\\
We end this work with various small comments.
\\~\\
\emph{\textbf{Shockwaves}}
\\~\\
The shockwave geometry in dS JT gravity is
\begin{equation}
ds^2 = -\frac{4}{(1-u v)^2} \, du \, dv - 4 \gamma \, \delta(u) \, du^2\,\label{c1}
\end{equation}
There is some finite window in which the right static observer can receive signals from the left static observer and vice versa, therefore early perturbations create traversability in dS. These shockwave configurations are discussed nicely in \cite{Aalsma:2020aib}, therefore we can skip the details.

For reasons explained before, shockwave calculations are not reliable in this context, but the elementary concepts of shockwaves still provide good intuition. 

We still expect that early perturbations create shockwaves, which have a certain strength. Observables sensitive to a shock have a response to the shock proportional to this strength. This strength grows like $\sim \exp(2\pi t/\beta)$ hence the universality of Lyapunov growth in gravity. Comparing our two calculations \eqref{f11} and \eqref{ftwoans} suggests the perturbation $\varphi_2(-t)$ creates some shockwave of strength proportional to
\begin{equation}
    \gamma \sim \frac{1}{\varepsilon_2\ln \varepsilon_2}\,G\ell \exp(2\pi t/\beta).\label{47}
\end{equation}
This matches with the predictions from a semiclassical shockwave analysis (see appendix \ref{app:shock}).
\\~\\
\emph{\textbf{Exact calculations}}
\\~\\
The identification of \eqref{bubo} as product of boundary bilocals can be used to compute bulk observables in dS JT gravity exactly \cite{blommaert2019clocks}. Schwarzian path integrals of products of bilocals are known \cite{Mertens:2017mtv,Blommaert:2018oro,Blommaert:2018iqz,Bagrets:2016cdf,Bagrets:2017pwq,Mertens:2018fds,Yang:2018gdb,Iliesiu:2019xuh,Mertens:2020pfe,Kitaev:2018wpr}, moreover the relation with AdS disk calculations suggests one possible prescription to include higher genus effects, which are also analytically tractable \cite{saad2019late,Blommaert:2020seb,blommaert2019clocks}. The motivations for this prescription in dS quantum gravity however, can be questioned. Better such motivations might fuel progress on the puzzles raised in \cite{Dyson:2002nt}.
\\~\\
\emph{\textbf{Ordering ambiguities}}
\\~\\
Consider two operators placed deep inside the AdS bulk, such that their lightcones intersect
\begin{equation}
    \raisebox{-10mm}{\includegraphics[width=46mm]{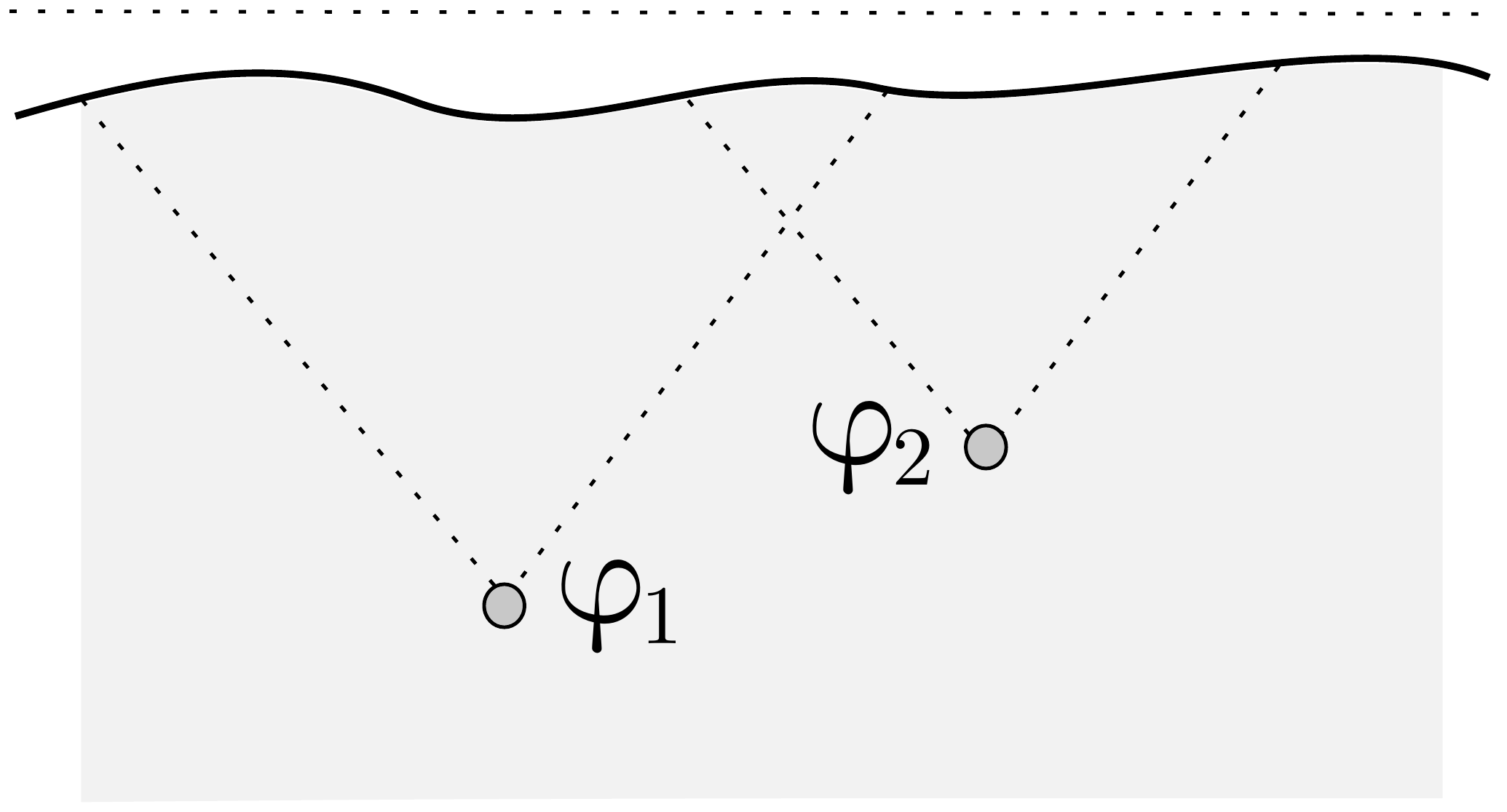}}
\end{equation}
Suppose one wants to calculate some out of time ordered correlator
\begin{equation}
    \average{\varphi_1(z_1,t_1)\varphi_2(z_2,t_2)\varphi_1(z_1,t_1)\varphi_2(z_2,t_2)}\,,
\end{equation}
then we end up computing the Schwarzian eight point function in \eqref{adspert}.

However here it is not obvious how to order the eight points on the Euclidean boundary, from which one should analytically continue to Lorentzian times, put differently, the correct sheet is less obvious. This emphasizes that operator ordering confusions are not an artifact of this dS context, but appear when dealing with bulk operators in quantum gravity \cite{blommaert2019clocks,Blommaert:2020yeo}.
%--------------------------------------------------
\subsection*{Acknowledgments}
%--------------------------------------------------
I thank Jordan Cotler for initial collaboration and for countless key discussions. Furthermore thanks to Kristan Jensen, Thomas Mertens and Douglas Stanford for useful discussions.

I gratefully acknowledge funding by the Belgian American Educational Foundation, and by the Stanford Institute for Theoretical Physics.

\begin{appendix}

%%%%%%%%%%%%%%%%%%%%%%%%%%%%%
% SECTION %
%%%%%%%%%%%%%%%
\section{Perturbative calculations}\label{app:fourpoint}
We review the piecewise analytic answer for the perturbative Schwarzian four point functions $\average{L(x_1,x_2)L(x_3,x_4)}$ \cite{Sarosi:2017ykf,Maldacena:2016upp}. Depending on the ordering of these four points on the Euclidean boundary, one distinguishes three disconnected patches on the principle sheet.

Inequivalent orderings are classified by topologically inequivalent line diagrams, obtained by connecting these boundary points $x_1$ to $x_2$ and $x_3$ to $x_4$ via lines which traverse the bulk. These inequivalent orderings or patches are
\begin{enumerate}
    \item Time ordered patch $x_1>x_2>x_3>x_4$
    \begin{align}
        \frac{\pi^2}{G\ell}\average{L(x_1,x_2)L(x_3,x_4)}=\left(\frac{x_1-x_2}{\tan \left(\frac{x_1-x_2}{2}\right)}-2\right)\left(\frac{x_3-x_4}{\tan \left(\frac{x_3-x_4}{2}\right)}-2\right).\label{exp1}
    \end{align}
    \item Nested patch $x_3>x_1>x_2>x_4$
    \begin{align}
        \frac{\pi^2}{G\ell}\average{L(x_1,x_2)L(x_3,x_4)}=\left(\frac{x_1-x_2}{\tan \left(\frac{x_1-x_2}{2}\right)}-2\right)\left(\frac{2\pi-x_3+x_4}{\tan\left( \frac{2\pi-x_3+x_4}{2}\right)}-2\right).\label{exp2}
    \end{align}
\item Out of time ordered patch $x_1>x_3>x_2>x_4$
\begin{align}
    & \frac{\pi^2}{G\ell}\average{L(x_1,x_2)L(x_3,x_4)}=\left(\frac{x_1-x_2}{\tan \left(\frac{x_1-x_2}{2}\right)}-2\right)\left(\frac{x_3-x_4}{\tan \left(\frac{x_3-x_4}{2}\right)}-2\right)\nonumber \\&\quad +2\pi\frac{\sin \left(\frac{x_1-x_2+x_3-x_4}{2}\right)-\sin \left(\frac{x_1+x_2-x_3-x_4}{2}\right)}{\sin \left(\frac{x_1-x_2}{2}\right)\sin\left(\frac{x_3-x_4}{2}\right)} +2\pi\frac{x_2-x_3}{\tan \left(\frac{x_1-x_2}{2}\right)\tan\left(\frac{x_3-x_4}{2}\right)}.\label{otocll}
\end{align}
\end{enumerate}

It is elementary to derive \eqref{340} using these formulas, following the prescription explained in the main text, one finds several contributions that grow exponentially with time
\begin{align}
    \average{L(u_3,v_3)L(u_2,v_2)}&=\frac{2}{\pi}\,G\ell \exp(2\pi t/\beta)\\
    \average{L(u_2,v_2)L(u_1,v_3)}+\average{L(u_2,v_2)L(u_3,v_1)}&=\nonumber  \left(1+\frac{\varepsilon_2^2}{4}\right)\frac{2}{\pi}\,G\ell \exp(2\pi t/\beta)\\
    \average{L(u_3,v_3)L(u_4,v_2)}+\average{L(u_2,v_2)L(u_2,v_4)}&=\frac{2}{\pi}\,G\ell \exp(2\pi t/\beta)\nonumber\\
    \average{L(u_1,v_3)L(u_4,v_2)}+\average{L(u_3,v_1)L(u_2,v_4)}&=\left( 1+\frac{\varepsilon_2^2}{4}+\frac{\pi-2}{4}\,\varepsilon_1\,\varepsilon_2\right)\frac{2}{\pi}\,G\ell \exp(2\pi t/\beta)\,.\nonumber
\end{align}
Adding these with the appropriate signs which follow from \eqref{signs} one then reproduces \eqref{340}
\begin{equation}
    \average{L(u_1,u_3,v_1,v_3)L(u_2,u_4,v_2,v_4)}= \frac{\pi-2}{2\pi}\,\varepsilon_1\,\varepsilon_2\,G\ell \exp(2\pi t/\beta)\,.
\end{equation}
Any other results quoted in the main text can be reproduced in similarly elementary manner.

Exponential growth in dS comes from exponentially small denominators, whereas in AdS it arises when evaluating sines in the numerator of \eqref{otocll} on large imaginary arguments \cite{Maldacena:2016upp}.
%--------------------------------------------------
\section{Shockwaves}\label{app:shock}
%--------------------------------------------------
We consider the shockwave geometry created by inserting some semiclassical operator $\varphi_2(-t)$
\begin{equation}
ds^2 = -\frac{4}{(1-u v)^2} \, du \, dv - 4 \gamma \, \delta(u) \, du^2\,,\label{c1}
\end{equation}
and seek to determine how the strength $\gamma$ of the shock depends on $t$ and $\varepsilon_2$. Tracking overall constants is pointless, as emphasized in the main text. The following is based mostly on \cite{Roberts:2014ifa}. 
The shockwave is sourced by a stress tensor
\begin{equation}
    T_{u u}(u)=E\,\delta(u)\quad,\quad E=\int_{-\infty}^{+\infty}du\, T_{u u}(u)\,,
\end{equation}
General relativity fixes $\gamma \sim G\ell\,E$, therefore one can calculate the strength of the shockwave via the stress tensor in the state obtained by inserting this semiclassical operator $\varphi_2(-t)$
\begin{equation}
    \gamma\sim G\ell\,\int_{-\infty}^{+\infty}du\,\frac{\average{\varphi_2(-t)\, T_{u u}(u)\,\varphi_2(-t)}}{\average{\varphi_2(-t)\varphi_2(-t)}}\,.\label{c3}
\end{equation}
We can compute the denominator using the Klein-Gordon inner product
\begin{equation}
    \average{\varphi_2(-t)\varphi_2(-t)}\sim \int_{-\infty}^{+\infty} du\,\average{\varphi_2(-t)\varphi_2(u)}\partial_u \average{\varphi_2(-t)\varphi_2(u)}^*\,.\label{c4}
\end{equation}
Here
\begin{equation}
    \average{\varphi_2(-t)\varphi_2(u)}= \,_2F_1(\Delta,1-\Delta,1,z)\quad,\quad z=\frac{1+u \exp(2\pi t/\beta)}{2}\,,\label{c5}
\end{equation}
which approaching the singular point becomes
\begin{equation}
     \average{\varphi_2(-t)\varphi_2(u)}\sim \ln \frac{1-u\exp(2\pi t/\beta)}{2}\quad,\quad \partial_u  \average{\varphi_2(-t)\varphi_2(u)}\sim \frac{\exp(2\pi t/\beta)}{u\exp(2\pi t/\beta)-1}\,.\label{c7}
\end{equation}
The regulator moves the pole off the contour and the residue theorem applies, so we recover
\begin{equation}
     \average{\varphi_2(-t)\varphi_2(-t)}\sim \ln \varepsilon_2\,.
\end{equation}
Furthermore
\begin{equation}
    T_{u u}(u)\sim \,:\partial_u \varphi_2(u)\,\partial_u\varphi_2(u):\,,
\end{equation}
therefore we can compute the nominator of \eqref{c3} by taking Wick contractions
\begin{equation}
    \nn\int_{-\infty}^{+\infty} du \average{ \varphi_2(-t)\,T_{u u}(u,0)\,\varphi_2(-t)}\sim \int_{-\infty}^{+\infty} d u\,\partial_u  \average{\varphi_2(-t)\varphi_2(u)}\,\partial_u  \average{\varphi_2(-t)\varphi_2(u)}^*\,.
\end{equation}
The poles are slightly above and below the real axis, so elementary contour integration gives
\begin{equation}
    \average{\varphi_2(-t) E\,\varphi_2(-t)}\sim \frac{1}{\varepsilon_2}\,\exp(2\pi t/\beta)\,,
\end{equation}
therefore the strength of the shockwave goes like
\begin{equation}
    \gamma \sim \frac{1}{\varepsilon_2 \ln \varepsilon_2}\,G\ell \exp(2\pi t/\beta)\,.
\end{equation}
\end{appendix}

\bibliography{refs}
\bibliographystyle{JHEP}

\end{document}